\documentclass[aps,twocolumn,letterpaper,longbibliography]{revtex4-1}

\usepackage{amssymb}
\usepackage{amsmath}
\usepackage{graphicx}
\usepackage{extarrows}
\usepackage{url}
\usepackage{varioref}
\usepackage{hyperref}
\usepackage{cleveref}
\newcommand\ket[1]{\ensuremath{|#1\rangle}}

\usepackage{xcolor}
\usepackage[normalem]{ulem}

\usepackage{verbatim} 

\begin{document}

\title{Many-body Chemical Reactions in a Quantum Degenerate Gas} 

\author{Zhendong Zhang$^1$}
\author{Shu Nagata$^1$}
\author{Kaixuan Yao$^1$}
\author{Cheng Chin$^1$}

\affiliation{$^1$James Franck Institute, Enrico Fermi Institute and Department of Physics, University of Chicago, Chicago, Illinois 60637, USA}

\begin{abstract}
Chemical reactions in the quantum degenerate regime are described by mixing of matterwave fields. Quantum coherence and bosonic enhancement are two unique features of many-body reactions involving bosonic reactants and products. Such collective reactions of chemicals, dubbed ``super-chemistry'', is an elusive goal in quantum chemistry research. Here we report the observation of coherent and collective reactive coupling between Bose condensed atoms and molecules near a Feshbach resonance. Starting from an atomic condensate, the reaction begins with a rapid formation of molecules, followed by oscillations of their populations in the equilibration process. Faster oscillations are observed in samples with higher densities, indicating bosonic enhancement. We present a quantum field model which describes the dynamics well and identifies three-body recombination as the dominant reaction process. Our findings exemplify the highly sought-after quantum many-body chemistry and offer a new paradigm for the control of quantum chemical reactions.  
\end{abstract}

\maketitle 

Ultracold atoms and molecules form an ideal platform toward controlling chemical reactions at the level of single internal and external quantum states. Ultracold molecules can be prepared in an individual internal state by, e.g., magneto- ~\cite{Julienne2006_2} and photoassociation~\cite{Julienne2006_1} of ultracold atoms 
and direct laser cooling~\cite{Shuman2010}.
External motion of molecules can be constrained by loading them into optical lattices~\cite{Yan2013}
or tweezers~\cite{Cairncross2021}. These experiments lead to 
the realization of state-to-state ultracold chemistry~\cite{Joschka2017,Rui2017,Liu2021}.

A number of experiments on cold molecules have reached the regime of quantum degeneracy, which promise new forms of molecular quantum matter and reaction dynamics. For instance, molecular Bose-Einstein condensates (BECs)
formed in atomic Fermi gases have stimulated tremendous interest in the BEC-BCS (Bardeen-Cooper-Schrieffer) crossover~\cite{Qijin2005,Giorgini2008}. Degenerate fermionic molecules are created by magneto-association of bosonic and fermionic atoms and optical transitions to the lowest rovibrational state~\cite{Marco2019}. Here quantum degeneracy suppresses chemical reactions due to the fermion anti-bunching effect~\cite{Marco2019}.

\begin{figure}[h!]
    \centering
    \includegraphics[width=82mm]{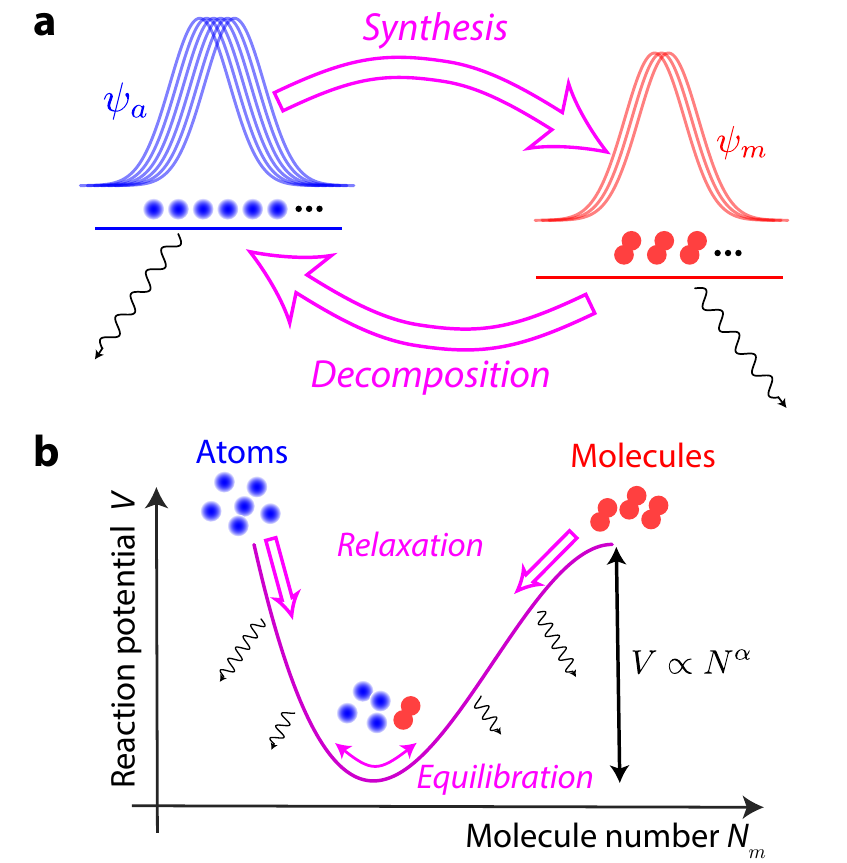}
    \caption{\textbf{Reactive coupling between atomic and molecular quantum fields.} \textbf{a}, Bose condensed atoms described by a single wavefunction $\psi_a$ are coupled to molecules condensed in the state $\psi_m$. The coupling synthesizes and decomposes molecules. Wavy lines represent dissipation. \textbf{b}, We introduce a reaction potential $V$ to describe the many-body dynamics of the atomic and molecular fields. A pure sample of atoms or molecules first relaxes towards lower potential and then equilibrates near the potential minimum. Due to bosonic stimulation, the potential scales as $V\propto N^\alpha$, where $N$ is the total particle number and $\alpha$ is the reaction order, see text.
    }
    \label{fig:Fig1}
\end{figure}

Recently, molecular BECs are realized based on atomic BECs near a Fesbach resonance~\cite{Zhang2021}. The reactive coupling between condensed atoms and molecules promises a new regime of quantum chemistry, dubbed ?quantum super-chemistry?, which highlights the coherent coupling of macroscopic matterwaves and Bose stimulation of the reaction process~\cite{Heinzen2000,Malla2021}. A key feature of the coherence is the collective oscillations between the reactant and product populations. Because of Bose statistics, enhancement of the reaction dynamics is anticipated to significantly modify the branching ratio~\cite{Moore2002}.

At quantum degeneracy, reaction dynamics come fundamentally from mixing of the matterwave fields of reactants and products. For instance, consider Feshbach coupling which converts two atoms into one molecule and \textit{vice versa}, 
described by the chemical equation: $\mathrm{A} + \mathrm{A}\longleftrightarrow\mathrm{A}_2$. In a quantum gas, the reaction is described by the many-body Hamiltonian with reaction order $\alpha=3$:
\begin{gather}
\hat{H} = \varepsilon_m\hat{\psi}_m^\dagger\hat{\psi}_m + g_2\hat{\psi}_m^\dagger\hat{\psi}_a^2+g_2\hat{\psi}_a^{\dagger2}\hat{\psi}_m,
\label{H}
\end{gather}
where $\hat{\psi}_a$ ($\hat{\psi}_m$) is the atomic (molecular) field operator, $g_2$ is the Feshbach coupling strength and $\varepsilon_m$ is the energy of one bare molecule relative to two bare atoms. 
Here we define the reaction order $\alpha$ as the maximum number of field operators in the reaction terms.

Another prominent example that couples ultracold atoms and molecules is three-body recombination, where three colliding atoms are converted into a diatomic molecule and another atom, and \textit{vice versa}. This process is described by the chemical equation $\mathrm{A}+\mathrm{A}+\mathrm{A}\longleftrightarrow\mathrm{A}_2+\mathrm{A}$. At quantum degeneracy, the recombination process can resonantly couple atomic and molecular fields as
\begin{gather}
\hat{h}_{\mathrm{3}} = g_3\hat{\psi}_m^\dagger\hat{\psi}_a^\dagger\hat{\psi}_a^3+g_3\hat{\psi}_a^{\dagger3}\hat{\psi}_a\hat{\psi}_m,
\label{H3}
\end{gather}
where $g_3$ is the recombination coupling strength. Here the reaction order is $\alpha = 5$. 

To understand the dynamics of the coupled quantum fields, we present the following picture. 
We show that the molecular population $\hat{N}_m = \hat{\psi}_m^\dagger\hat{\psi}_m$ follows the form of an ``energy conservation" law as
\begin{align}
    \frac{\hbar^2}{2}\dot{\hat{N}}_m^2 + \hat{V}(\hat{N}_m) = const., 
    \label{Conservation}
\end{align}
where $\hbar^2\dot{\hat{N}}_m^2/2$ resembles the kinetic energy and we introduce the many-body reaction potential $\hat{V} =[\hat{N}_m,\hat{H}]^2/2+const.$
(see Supplement). In this picture, the system tends towards lower potential. Quantum fluctuations of the nonlinear field coupling, however, can effectively damp the dynamics of the populations ~\cite{Vardi2001,Richter_2015}. In experiments, damping can also come from inelastic scattering and coupling to a thermal field. Thus one expects that the system first relaxes towards the potential minimum, and 
then equilibrates near the minimum with small amplitude coherent oscillations (see Fig.~\ref{fig:Fig1}). 
In the thermodynamic limit with total particle number $N\gg 1$, the reaction potential and the oscillation frequency near the minimum scale with the particle number as $V\propto N^\alpha$ and $\omega_0\propto N^{\alpha/2-1}$. The dependence on the particle number signals bosonic enhancement of the reaction dynamics~\cite{Heinzen2000,Vardi2001}.

In this paper, we report the observation of coherent and Bose stimulated reactions between Bose condensed Cs atoms and Cs$_2$ molecules. The reaction is initiated by tuning the magnetic field near a narrow $g-$wave Feshbach resonance, which couples scattering atoms and diatomic molecules in a single high-lying rovibrational state (See Supplement). Near the resonance, atomic and molecular populations quickly relax toward a dynamical equilibrium, followed by coherent oscillations between atoms and molecules in the equilibration process. We show that the oscillation frequency strongly depends on the particle number. From the dependence, we conclude that three-body recombination is the dominant reaction process that couples the atomic and molecular fields near the Feshbach resonance.

\begin{figure}
    \centering
    \includegraphics[width=82mm]{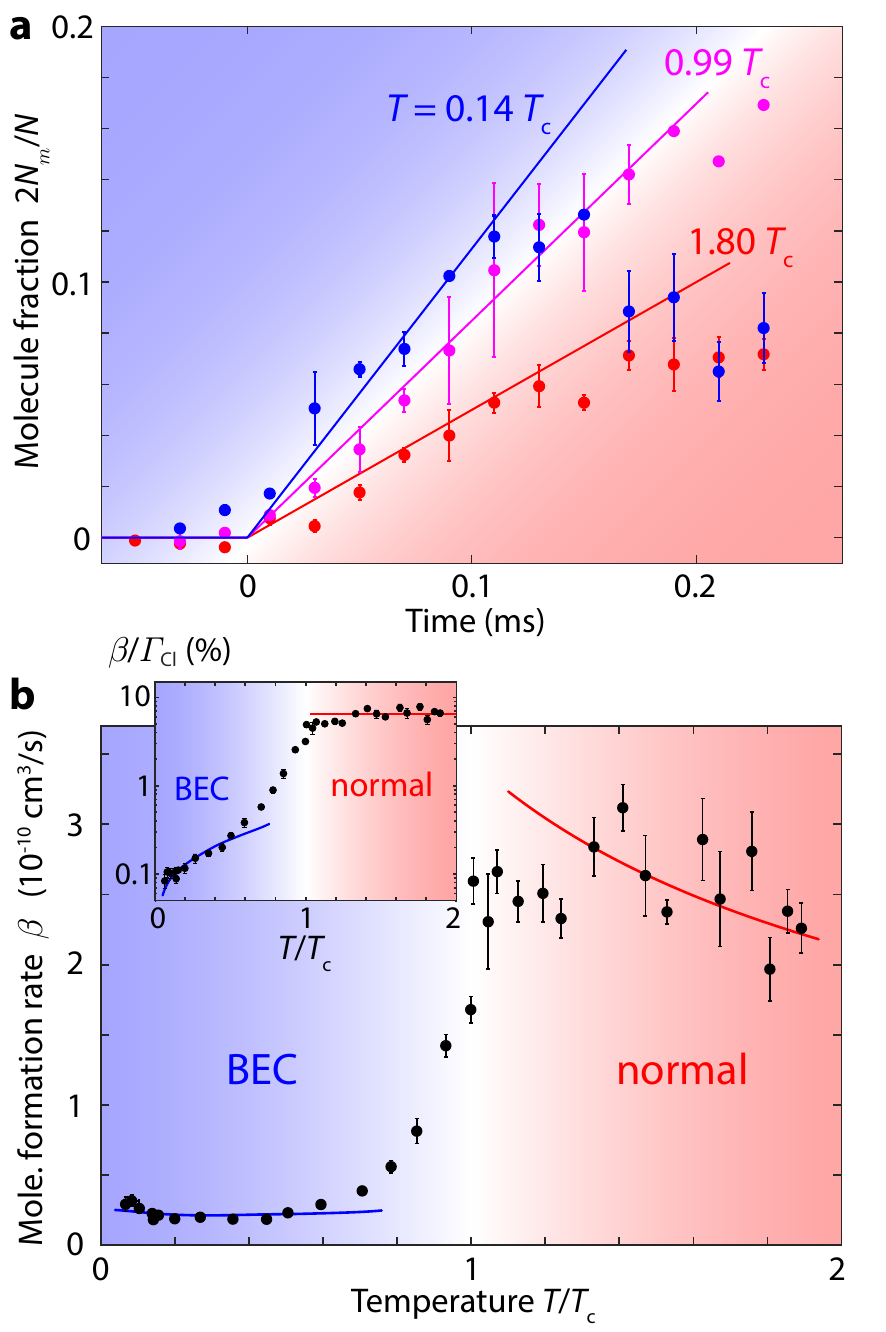}
    \caption{\textbf{Comparison of molecule formation rate in classical and quantum degenerate regimes.} \textbf{a},  Dynamics of molecule formation in an atomic gas after quenching the magnetic field  $3(2)$~mG above the Feshbach resonance at $B_0=19.849(2)$~G. The solid lines are fits to the data in the initial growth stage for extraction of the molecule formation rate $\dot{N}_{\mathrm{m}}$. \textbf{b}, The extracted molecule formation rate coefficient $\beta$ above and below the critical temperature $T_c$. The red line is a fit to the data based on the classical kinetic theory prediction $\beta = b_{\mathrm{Cl}}\Gamma_{\mathrm{Cl}}$, from which we obtain the classical branching ratio $b_{\mathrm{Cl}} = 7(1)\%$ (see text). The blue line fits the data in the quantum regime 
    with $\beta = b_{\mathrm{Q}}\Gamma_{\mathrm{Q}}$, which gives the quantum branching ratio $b_{\mathrm{Q}} = 3.9(3)\%$ (see text). The inset shows the rate coefficient normalized to the classical gas expectation $\Gamma_{\mathrm{Cl}}$. In panel \textbf{a} error bars represent one standard deviation of the mean, estimated from 4-8 measurements. In panel \textbf{b} error bars represent 95\% confidence intervals of the mean.}
    \label{fig:Fig2}
\end{figure}

\begin{figure*}
    \centering
    \includegraphics[width = 162mm]{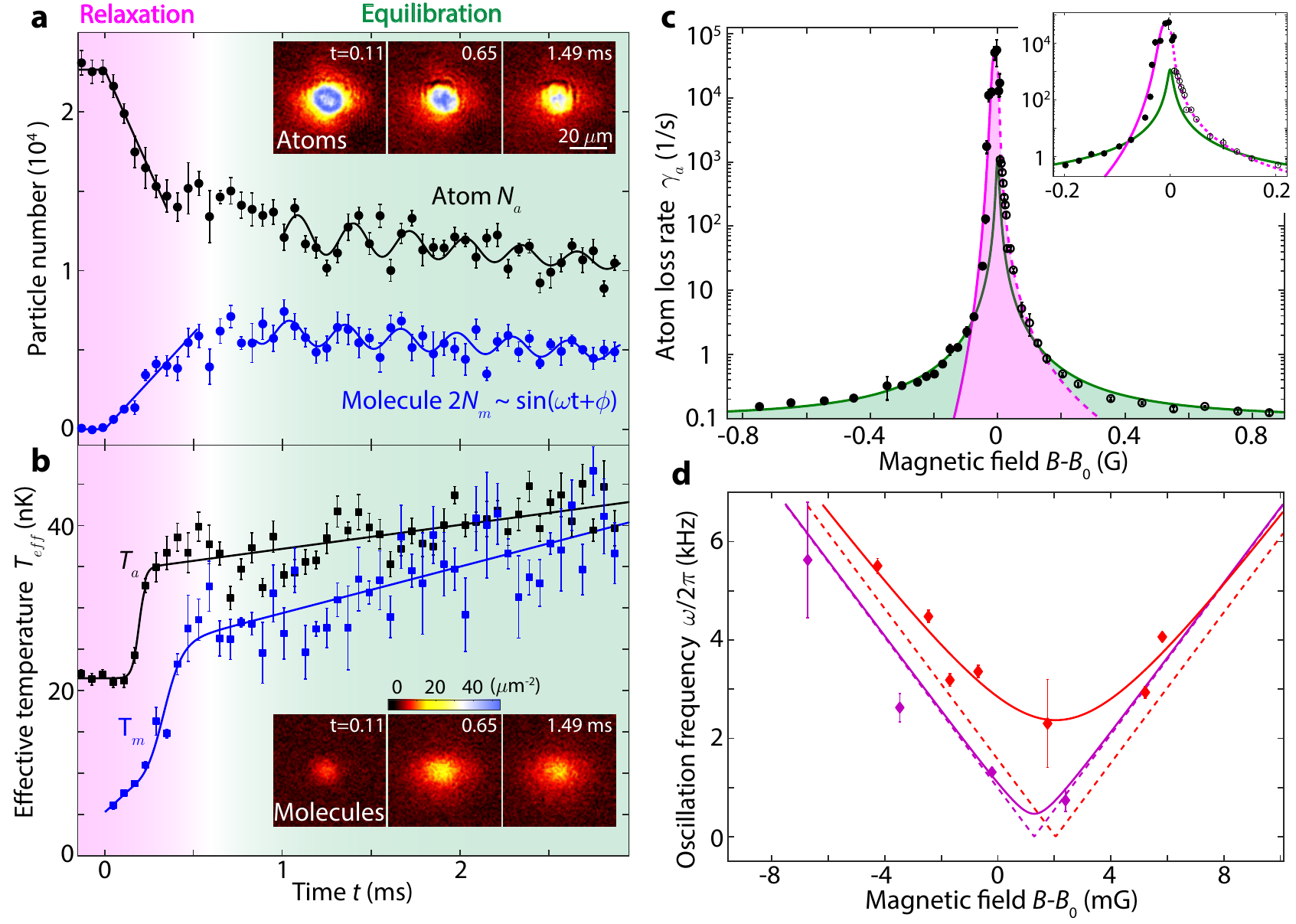}
    \caption{\textbf{Coherent reaction dynamics in quantum gases of atoms and molecules across a Feshbach resonance.} 
    \textbf{a}, Evolution of atomic and molecular populations in an atomic BEC after quenching the magnetic field $2(1)$~mG below the resonance. Solid lines are fits to capture the dynamics in relaxation and equilibration processes. \textbf{b}, Effective temperatures determined from time-of-flight measurements of the atoms and molecules at the same field. Solid lines are guides to the eye. Insets in panels \textbf{a} and \textbf{b} are sample images of atoms and molecules after the time-of-flight (see Supplement). \textbf{c}, Loss rate of atoms immediately after the quench. Solid (empty) circles represent samples prepared below (above) the resonance. Green line is a Lorentzian fit with center at $B_0$ and width $\Delta B$. Magenta solid (dashed) line is a fit near and below (above) the resonance based on $\gamma_a = \gamma_{0}/[1+|(B-B_0)/\delta B|^{\epsilon_{\pm}}]$ (see text), from which we obtain the exponent $\epsilon_{-} = 6(2)$ ($\epsilon_{+} = 2.9(4)$) below (above) the resonance. Inset is a zoomed-in view near the resonance. \textbf{d}, Oscillation frequency of molecular populations from atomic samples at mean BEC density of $2.9\times 10^{13}\mathrm{cm}^{-3}$ and BEC fraction of $80\%$ (red) and $60\%$ (purple). Solid lines are empirical fits based on $\omega = \sqrt{\delta\mu^2(B-B_m)^2+\omega_0^2}$, where $B_m$ and $\omega_0$ are fitting parameters. The values of $B_m$ from the fits are consistent with the resonance position $B_0$ within our measurement uncertainty. Dashed lines are the asymptotes $\omega = |\delta\mu(B-B_m)|$. Data in panels \textbf{a} and \textbf{b} are averages of 3-4 measurements, and error bars represent one standard deviation of the mean. Data in panels \textbf{c} and \textbf{d} are obtained from the fits, see Supplement, and error bars represent 95\% confidence intervals. }
    \label{fig:Fig3}
\end{figure*}

Our experiment starts with an ultracold Bose gas of $6\times10^3$ to $5\times10^5$ 
cesium atoms in an optical trap. The atoms can form a pure BEC either in a three-dimensional (3D) harmonic potential or a 2D square well potential~\cite{Clark2017}.
We induce the reaction by switching the magnetic field near the \textit{g}-wave Feshbach resonance, which can convert an atomic BEC into a molecular BEC~\cite{Zhang2021}.  
We determine the resonance position $B_0 = 19.849(2)$~G, resonance width $\Delta B = 8.3(5)$ mG, and the relative magnetic moment $\delta\mu = h\times0.76(3)$~MHz/G, where $h$ is the Planck constant, from measurements of the molecular binding energy $\varepsilon_m \approx \delta\mu(B-B_0)$ and the scattering length (see Supplement). After the reaction, we decouple the atoms and molecules by quickly tuning the magnetic field far off the resonance and image each independently~\cite{Zhang2021}. 

To show that chemical reactions follow different rules in a degenerate quantum gas versus in a normal gas, we compare the molecule production rate for samples prepared above and below the BEC critical temperature $T_c$. We extract the molecule production rate coefficient $\beta = \dot{N}_m/N_0n_0$ right after the magnetic field switch, where $N_0$ and $n_0$ are the initial total atom number and mean atomic density, respectively, see Fig.~\ref{fig:Fig2}.

The measured molecule formation rate shows distinct behaviour in the two regimes. In a thermal gas with temperature $T>T_c$, the molecule formation rate is $\beta = b_{\mathrm{Cl}}\Gamma_{\mathrm{Cl}}$, where $\Gamma_{\mathrm{Cl}}$ and $b_{\mathrm{Cl}}$ are the classical atomic collision rate coefficient and the branching ratio into the molecular state, respectively. 
Near the resonance, the collision rate coefficient is unitarity limited as $\Gamma_{\mathrm{Cl}} = 8h^2(\pi^3m^3k_{\mathrm{B}}T)^{-1/2}$~\cite{Chin2010}
, where $m$ is the atomic mass and $k_{\mathrm{B}}$ is the Boltzmann constant. Our measurement in the thermal regime is consistent with the $T^{-1/2}$ scaling. From the fit we extract the branching ratio $b_{\mathrm{Cl}} = 7(1)\%$ (see Fig.~\ref{fig:Fig2}b). 

\begin{figure*}
    \centering
    \includegraphics[width = 170mm]{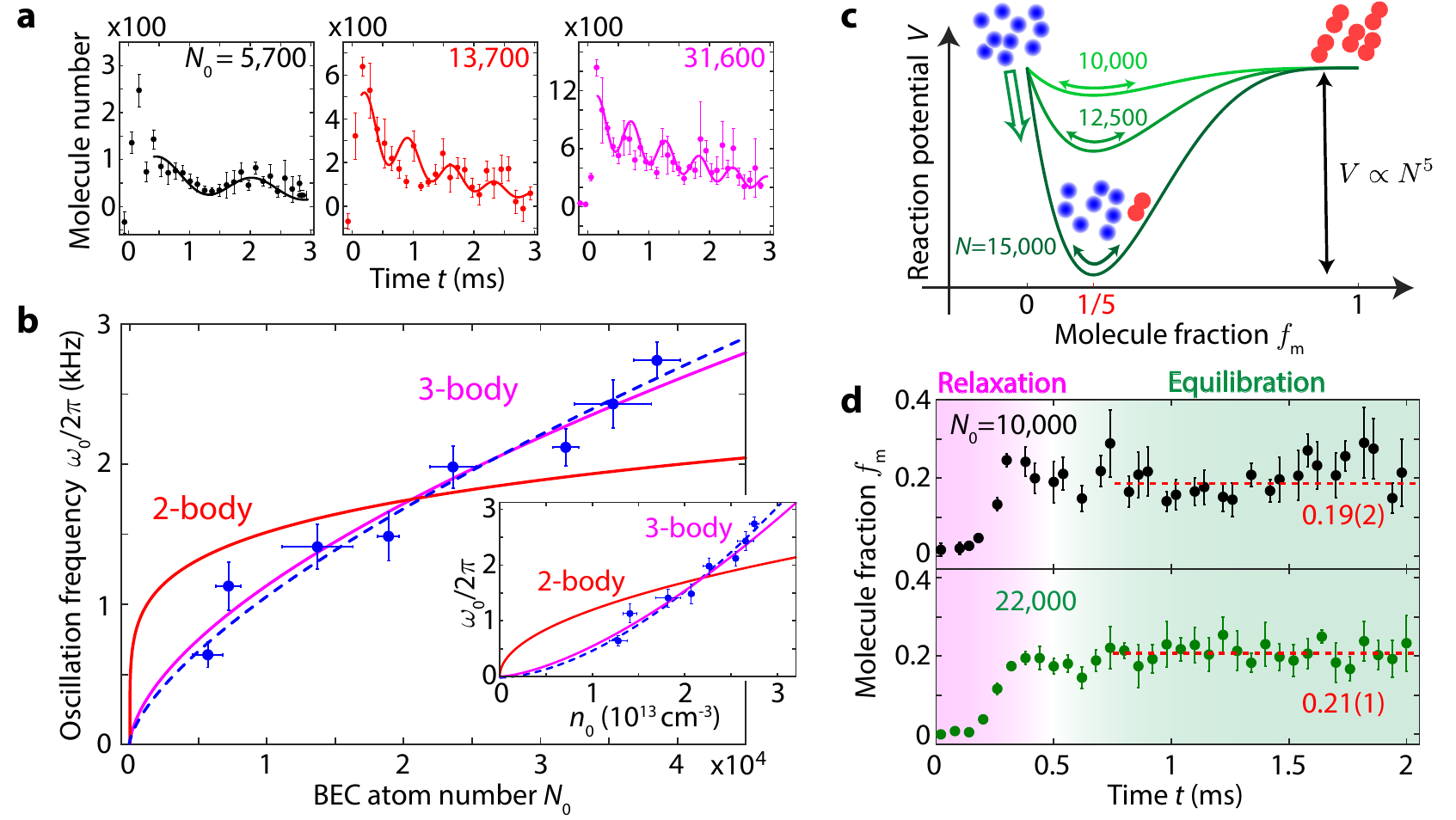}
    \caption{\textbf{Bose-enhanced atom-molecule reaction dynamics on Feshbach resonance.} \textbf{a}, Molecules formed in atomic BECs with different initial atom numbers $N_0$ following the magnetic field quench to $1(1)$~mG below the Feshbach resonance. 
    Solid lines are fits to the data. \textbf{b}, The extracted oscillation frequencies for different initial atom number $N_0$. The red (magenta) solid line is a power law fit with exponent given by the two-(three-)body model. The blue dashed line is a power law fit with a free exponent, which yields the scaling $\omega_0\propto N_0^{0.7(2)}$. In the inset, we show the frequency dependence on the mean atomic density $n_0$ and the associated fits that yields the scaling $\omega_0\propto n_0^{1.7(4)}$. \textbf{c}, The reaction potential $V\equiv V_3 \approx - 4g_3^2N^5f_m(1-f_m)^4$ of the three-body process described in Eq.~(\ref{H3}) for different total particle number $N$ (green solid lines). In the thermodynamic limit $N\gg 1$, the minimum occurs at molecule fraction $f_m \equiv 2N_m/N = 1/5$. \textbf{d}, Evolution of molecule fraction $f_m$ for different initial atom number $N_0$. The mean molecule fractions in the equilibration phase $f_m = 19(2)\%$ and $f_m = 21(1)\%$ are consistent with the predicted minimum position of $V_3$ at $f_m = 1/5$. Here the uncertainties represent $95\%$ confidence intervals. Data in panels \textbf{a} and \textbf{d} are averages of 3-5 measurements and error bars represent one standard deviation of the mean. Error bars in panel \textbf{b} represent 95\% confidence intervals. }
    \label{fig:Fig4}
\end{figure*}


Entering the quantum degenerate regime $T<T_c$, we observe a steep drop in the rate coefficient, see Fig.~\ref{fig:Fig2}b. 
At low temperatures, we model the rate coefficient as $\beta = b_{\mathrm{Q}}\Gamma_{\mathrm{Q}}$, where $b_{\mathrm{Q}}$ and $\Gamma_{\mathrm{Q}} = 4\sqrt{3}(h/m)(6\pi^5n)^{-1/3}$ are the branching ratio and the rate coefficient predicted by the universal theory in the quantum regime  ~\cite{Jin2014,Eismann2016}. The model fits the measurement well, and we extract the branching ratio to be $b_{\mathrm{Q}} = 3.9(3)\%$. 
The sharp transition of molecule formation rate around the critical temperature $T_c$ indicates different laws in the classical and quantum degenerate regimes. 

Close examination of molecule formation dynamics in atomic BECs reveals additional interesting features of quantum many-body reactions. To understand the underlying reaction processes, we study the atom loss rate $\gamma_a = -\dot{N}_a/N_0$ right after switching the magnetic field, where $N_a$ is the atom number (see Fig.~\ref{fig:Fig3}c). 
Far from the resonance $|B-B_0|\gg\Delta B$, atoms decay slowly and the loss rate follows a symmetric Lorentzian profile centered at the resonance $\gamma_a\propto (B-B_0)^{-2}$. We attribute the lineshape to the Feshbach coupling ~\cite{Chin2010}. 

Near the resonance, the loss rate greatly exceeds the expectation from the Lorentzian profile. This rapid atom loss only lasts for a few 100~$\mu$s and is accompanied with fast molecule production and heating of both atoms and molecules. We identify this fast process as the relaxation dynamics described in Fig.~\ref{fig:Fig1}b. In order to characterize the enhanced reaction rate, we fit the loss rate near the resonance as $\gamma_a\propto [1+|(B-B_0)/\delta B|^{\epsilon_{\pm}}]^{-1}$ from which we extract the exponents $\epsilon_+ = 2.9(4)$ above the resonance and $\epsilon_- = 6(2)$ below the resonance. The exponents $\epsilon_{\pm}$ larger than 2 are consistent with the enhanced atom loss near the resonance beyond the Lorentzian profile. 

The relaxation dynamics stem from three-body recombination, evidenced by the fast heating of both species in the relaxation phase, see Fig.~\ref{fig:Fig3}b~\cite{Greene2017}. In addition, the measured exponent $\epsilon_{+} = 2.9(4)$ from the enhanced atom loss is consistent with the predicted value of $3.5$ for three-body recombination near a narrow Feshbach resonance~\cite{Petrov2004}. We attribute the even larger exponent $\epsilon_{-} = 6(2)$ below the resonance to bosonic enhancement of the three-body process. 

Following the relaxation, both atomic and molecular populations oscillate for several ms before they slowly decay over a much longer time scale (see Fig.~\ref{fig:Fig3}a,b and Supplement). The oscillation is consistent with the equilibration dynamics near the reaction potential minimum described in Fig.~\ref{fig:Fig1}b. 
The frequency $\omega$ of the oscillation depends on the magnetic field and is well fit to $\omega = \sqrt{\varepsilon_m^2/\hbar^2+\omega_0^2}$~(see Fig.~\ref{fig:Fig3}d). Far from the resonance, the frequency approaches the molecular binding energy $|\varepsilon_m|$. On resonance with $\varepsilon_m = 0$, the frequency $\omega = \omega_0$ is given by the collective reactive coupling between the atomic and molecular fields. Large-amplitude oscillations are also observed in samples with magnetic field modulation, see Supplement.

To demonstrate the many-body nature of the reactive coupling, we probe the atom-molecule oscillations right on the Feshbach resonance with different initial atom number $N_0$ and mean density $n_0$. 
After quenching the magnetic field, we observe that samples with higher populations and densities display faster oscillations, see Fig.~\ref{fig:Fig4}a. Fitting the data, we obtain the scaling $\omega_0\propto N_0^{0.7(2)}$ and $\omega_0\propto n_0^{1.7(4)}$, 
see Fig.~\ref{fig:Fig4}b. 
Note that the two scalings are linked by $n_0\propto N_0^{2/5}$ for BECs in a harmonic trap~\cite{Pethick2008}. The particle number dependence of the reactive coupling supports bosonic enhancement of the reaction process.

The scaling with respect to the particle number also reveals the underlying reaction mechanism. 
For the two-body process described in Eq.~(\ref{H}), we derive the effective potential $V_2 = -g_2^2N^3f_m(1-f_m)^2+O(N^2)$, where $f_m = 2N_m/N$ is the molecule fraction, from which the resonant oscillation frequency is calculated to be $\omega_0\propto N^{1/5}$ in a harmonic trap (see Supplement). For the three-body recombination process described in Eq.~(\ref{H3}), the effective potential is $V_3 = -g_3^2N^5f_m(1-f_m)^4+O(N^4)$, which yields the scaling $\omega_0\propto N^{3/5}$. Our measurement agrees well with the three-body model, see Fig.~\ref{fig:Fig4}b. 

Moreover, we find the molecule fraction oscillates around 20(1)\% in the equilibration phase, which is consistent with the minimum position of the reaction potential $V_3$ at $f_m = 1/5$
~(see Fig.~\ref{fig:Fig4}c,d). Two-body Feshbach process, on the other hand, predicts a different minimum of $V_2$ at $f_m = 1/3$. 

To conclude, we observe collective many-body chemical reactions in an atomic BEC near a Feshbach resonance. The dynamics are well described by a quantum field model derived from three-body recombination. In particular, the coherent oscillations of atomic and molecular fields in the equilibration phase support quantum coherence and Bose enhancement of the reaction process. The observation of coherent and collective chemical reactions in the quantum degenerate regime paves the way to explore the interplay between many-body physics and ultracold chemistry.\\


\noindent\textbf{Acknowledgement}
\noindent We thank P. Julienne, K. Levin, D. Mazziotti, D. DeMille and K.-K. Ni for helpful discussions. We thank K. Patel and L. Weiss for carefully reading the manuscript. We thank J. Jachinowski for experimental assistance and carefully reading the manuscript. This work was supported by
the National Science Foundation under Grant No. PHY1511696 and PHY-2103542, by the Air Force Office of Scientific Research under award number FA9550-21-1-0447. Z.Z. is supported by the Grainger Graduate Fellowship. S.N. acknowledges support from the Takenaka Scholarship Foundation.


\clearpage
\widetext
\setcounter{equation}{0}
\setcounter{figure}{0}
\setcounter{table}{0}
\setcounter{page}{1}
\makeatletter
\renewcommand{\theequation}{S\arabic{equation}}
\renewcommand{\thefigure}{S\arabic{figure}}
\renewcommand{\thetable}{S\arabic{table}}

\noindent \textbf{Supplementary Material}\\

\section{Experimental Procedure}
Our experiment starts with a ultracold Bose gas of 6,000 to 470,000 $^{133}$Cs atoms at a temperature of 2 to 232 nK in a 3D harmonic trap. We tune the temperature and atom number by changing the trap depth at the end of evaporation process~\cite{Hung2008}. The harmonic trap frequencies are ($\omega_x$,$\omega_y$,$\omega_z$) = $2\pi\times$(24,13,74) to $2\pi\times$(36,15,91) Hz. The atoms are polarized into the hyperfine ground state $\ket{F=3,m_{\mathrm{F}}=3}$, where $F$ and $m_{\mathrm{F}}$ are quantum numbers for the total spin and its projection along the magnetic field direction, respectively. The narrow g-wave Feshbach resonance couples $\mathrm{Cs}$ atoms into $\mathrm{Cs}_2$ molecules at $\ket{f=4,m_f=4;l=4,m_l=2}$, where $f$ and $l$ represent quantum numbers for the sum of the spins of two individual atoms and the orbital angular momentum of a molecule, $m_f$ and $m_l$ are projections of $f$ and $l$ along the magnetic field direction~\cite{Chin2005}.

To induce the molecule formation dynamics, we quench the magnetic field close to the resonance position $B_0$ from 19.5 G where the samples are prepared. After holding for variable times, we switch the field back to either 19.5 G or 17.17 G to decouple atoms and molecules. We can image the remaining atoms at this field by absorption imaging. We can also wait for the remaining atoms to fly away after a resonant light pulse and image the molecules by jumping the field up to 20.4 G to dissociate them into atoms and then image the atoms from the dissociation~\cite{Zhang2021}. For the atom loss measurements shown in Fig.~\ref{fig:Fig3}c, BECs with $\sim$40,000 atoms are transferred from the harmonic trap to a 2D flat-bottomed optical potential before we quench the field to different values near the resonance~\cite{Clark2017}. For the rest of the data shown in Figs.~\ref{fig:Fig2}-\ref{fig:Fig4}, we start from atomic samples in the 3D harmonic dipole trap.

To measure the temperature of atoms or molecules (e.g. as is shown in Fig.~\ref{fig:Fig3}b), we release them into a horizontally isotropic harmonic trap for a quarter of the trap period, which converts the particle distribution from the real space to the momentum space~\cite{zhang2022coherent}. We extract the temperature $T$ by fitting the momentum distribution with the condensate around zero momentum excluded using the Gaussian function $n(k_r) = n(0)\exp{[-\hbar^2k_r^2/(2mk_{\mathrm{B}}T)]}$, where $k_r$ is the radial wave number and $k_{\mathrm{B}}$ is the Boltzmann constant.

\section{Determination of the Feshbach resonance position and width}

To determine the position of the narrow g-wave Feshbach resonance in our system, we perform measurements of molecular binding energy at different offset magnetic fields using magnetic field modulation spectroscopy~\cite{Thompson2005,Lange2009} and find the field value where the binding energy reaches zero. 

\begin{figure}[h!]
    \centering
    \includegraphics[width=170mm]{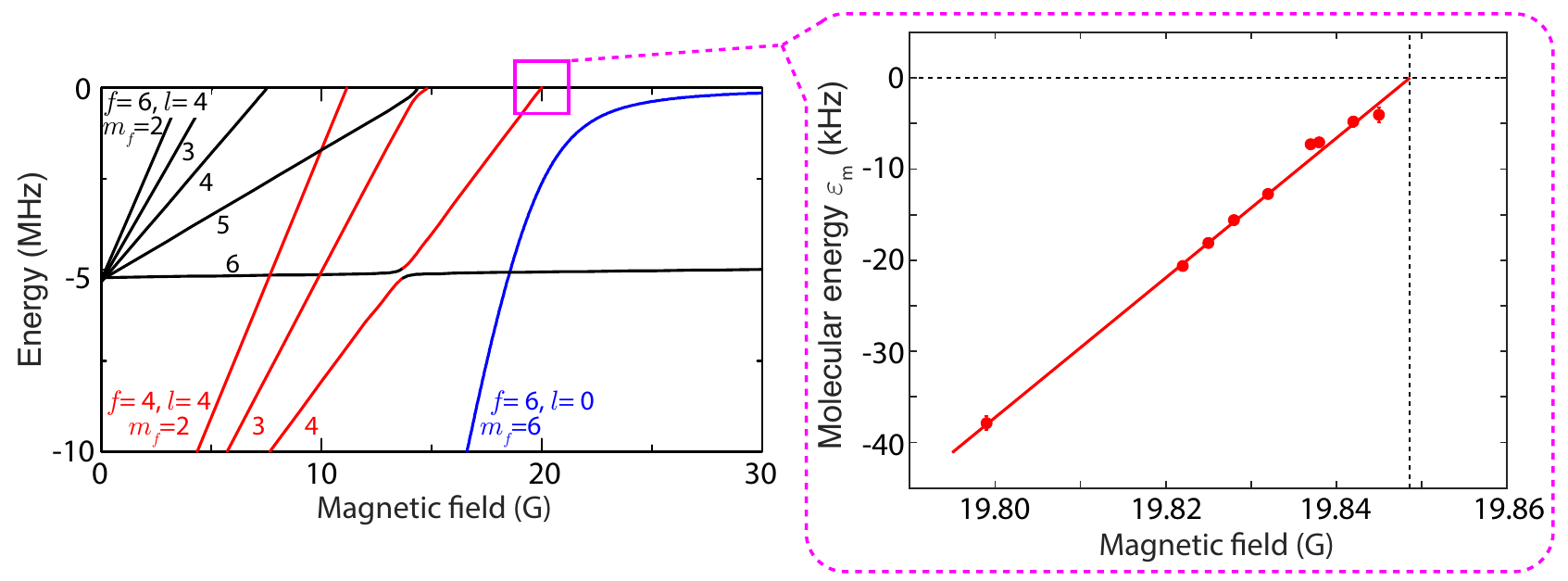}
    \caption{\textbf{Bound state energy diagram for cesium atoms in the hyperfine ground state $\ket{F=3,m_{\mathrm{F}}=3}$ and molecular energy measurement near the g-wave Feshbach resonance around 20 G using modulation spectroscopy.} \textbf{a}, Energy diagram for $\mathrm{Cs}_2$ molecular states close to the atomic scattering continuum adapted from Fig.~22 in Ref.~\cite{Chin2010}. \textbf{b}, Molecular energy $\varepsilon_\mathrm{m}$ obtained from modulation spectroscopy at different offset magnetic fields. The solid line is a linear fit which reaches 0 at $B_0 = 19.849(1)$ G.}
    \label{fig:FigS1}
\end{figure}

We start with atomic BECs at 23 nK prepared at 19.5 G. Then we quench the field to an offset value $B_{\mathrm{dc}}$ near the resonance and simultaneously modulate the field sinusoidally with an amplitude $B_{\mathrm{ac}} = 5$ mG for 5 ms. We scan the modulation frequency and measure the spectrum of the remaining atom number. From the atom loss peak of the spectrum due to the conversion from atoms into molecules, we extract the resonant frequency that corresponds to molecular energy at an offset magnetic field $B_{\mathrm{dc}}$ near the g-wave Feshbach resonance~\cite{zhang2022coherent, Thompson2005}, see Fig.~\ref{fig:FigS1}. We have confirmed that the resulting atom loss peak position is not sensitive to the modulation amplitude and modulation time. A linear fit to the data in Fig.~\ref{fig:FigS1} gives the resonance position $B_0 = 19.849(1)$ G where the molecular energy goes to zero. The slope of the linear fit gives the magnetic moment difference between two bare atoms and one bare molecule as $\delta\mu = h\times 0.76(3)$ MHz/G, which is consistent with Ref.~\cite{Chin2005}. We emphasize that for the narrow resonance we are using, the molecular energy approaches zero quadratically only within a small fraction of the resonance width. Our linear fit to the molecular energy data underestimates the resonance position by $\sim 0.3$ mG based on our calculation using the resonance width from the following scattering length measurements~\cite{Chin2010}. The systematic error of our calibration of the absolute magnetic field is less than 20 mG. 
Throughout this work, we perform the magnetic field calibration based on the same procedure to ensure a constant systematic error.

Next we measure the s-wave scattering length near the resonance to obtain the resonance width. Here the scattering length is inferred from the expansion of a quasi-2D BEC prepared with trap frequencies $(\omega_x,\omega_y,\omega_z) = 2\pi\times (11,13,895)$ Hz. During the expansion, the mean field interaction energy is converted into kinetic energy. We first prepare the BEC at an initial magnetic field $B_i = 20.481$ G or $19.498$ G where the scattering is $a_i$. The column density distribution of atoms in the Thomas-Fermi regime is~\cite{Hung2011}:
\begin{align}
    n(x,y) = [\mu - \frac{1}{2}m\omega_x^2x^2 - \frac{1}{2}m\omega_y^2y^2]/g_{\mathrm{2D}},
\end{align}
where $g_{\mathrm{2D}} = (\hbar^2/m)\sqrt{8\pi}a_i/l_z$ is the coupling strength, $l_z = \sqrt{\hbar/m\omega_z}$ is the harmonic oscillator length in the tightly confined z direction and $\mu = \sqrt{g_{\mathrm{2D}}Nm\omega_x\omega_y/\pi}$ is the chemical potential determined by $g_{\mathrm{2D}}$, the total atom number $N$ and the initial trap frequencies $\omega_x$ and $\omega_y$ in the horizontal plane.
\begin{figure}
    \centering
    \includegraphics[width=170mm]{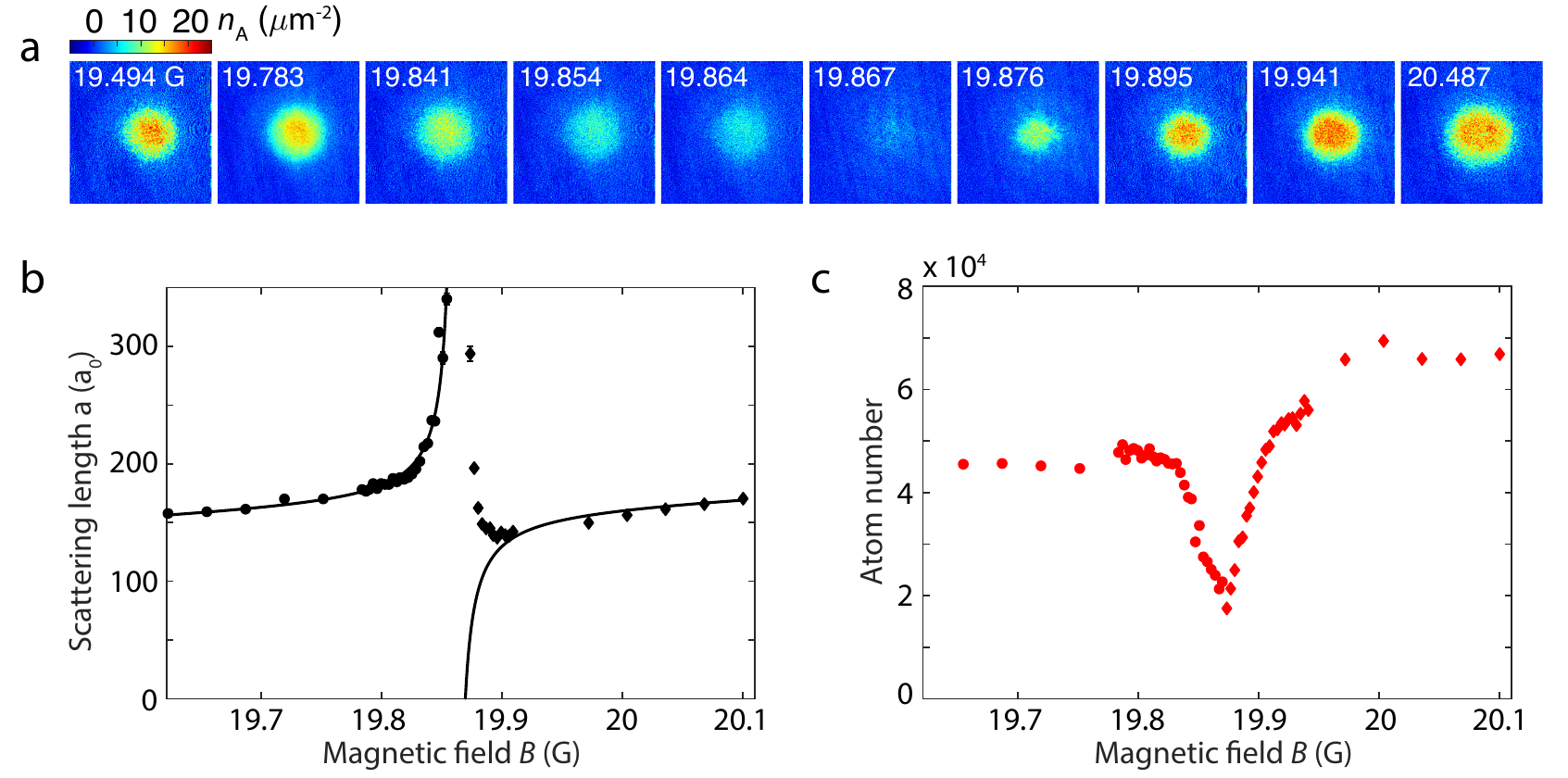}
    \caption{\textbf{Scattering length measurement near the narrow g-wave Feshbach resonance by time-of-flight.} \textbf{a}, Atomic density distributions after 20 ms time-of-flight at different magnetic fields near the Feshbach resonance. The images with $B<19.865$~G ($B>19.865$~G) come from initial BECs prepared below (above) the Feshbach resonance. \textbf{b}, Scattering length extracted from the Thomas-Fermi radii in the time-of-flight images, see text. The circular (diamond) data points come from initial BECs prepared below (above) the resonance. The solid line is a fit to the data excluding the points at $19.858~\mathrm{G}<B<19.909~\mathrm{G}$ based on Eq.~(\ref{aB_measurement}), from which we obtain the resonance width $\Delta B = 8.3(5)$ mG. The points at $19.855~\mathrm{G}<B<19.909~\mathrm{G}$ are excluded because of the heating effect near the resonance. \textbf{c}, Total atom number extracted from the time-of-flight images.}
    \label{fig:FigS2}
\end{figure}
Then we quench the magnetic field to a different value $B_f$ where the scattering length is $a_f$ and simultaneously switch off the harmonic trap in the horizontal plane. According to Ref.~\cite{Castin1996}, the dynamics of a BEC after the release follow a simple dilation with scaling parameters $\lambda_x(t)$ and $\lambda_y(t)$, which determine the density distribution at time $t$ as:
\begin{align}
   n(x,y,t) = \frac{\mu - \frac{1}{2}m\omega_x^2x^2/\lambda_x^2(t) - \frac{1}{2}m\omega_y^2y^2/\lambda_y^2(t)}{g_{\mathrm{2D}}\lambda_x(t)\lambda_y(t)}, 
\end{align}
where the scaling parameters evolve according to:
\begin{align}
\ddot{\lambda}_x(t) &= \frac{a_f}{a_i}\frac{\omega_x^2}{\lambda_x^2(t)\lambda_y(t)}\nonumber\\
\ddot{\lambda}_y(t) &= \frac{a_f}{a_i}\frac{\omega_y^2}{\lambda_y^2(t)\lambda_x(t)}.
\label{lambdas}
\end{align}

We scan the magnetic field and measure the Thomas-Fermi radii $R_j = \sqrt{2\mu\lambda_j^2(t)/m\omega^2_j}$ where $j = x,y$ after 20 ms expansion. Eventually we extract $a_f$ based on its one-to-one correspondence to the Thomas-Fermi radii according to Eq.~(\ref{lambdas}). The results are summarized in Fig.~\ref{fig:FigS2} and we fit the scattering length data using the formula~\cite{Chin2010}
\begin{align}
    a(B) = a_{\mathrm{bg}}[1+\eta(B-B_0)](1-\frac{\Delta B}{B-B_0}),
    \label{aB_measurement}
\end{align}
where we obtain the 
resonance width $\Delta B = 8.3(5)$ mG, the resonance position $B_0 = 19.861(1)$ G, the background scattering  length on resonance $a_{\mathrm{bg}} = 163(1)a_0$ and the slope of the background scattering length $\eta = 0.31(2)$/G. The background scattering length $a_{\mathrm{bg}}$ and the slope $\eta$ are consistent with Ref.~\cite{Berninger2013} and the resonance width $\Delta B$ is consistent with Ref.~\cite{Mark2018}, where a different method is used. The fitted resonance position deviates from that in the binding energy measurement by $\sim$10~mG, which we attribute to the heating of atoms near the resonance. The binding energy measurement, however, suffers less from the heating issue~\cite{Thompson2005}. Throughout the whole paper, we adopt the resonance position $B_0 = 19.849(1)$ G from the binding energy measurement.

\section{Many-body chemical reactions in quantum degenerate regime}

Here we derive the effective potential $V$ for three-body and two-body processes and calculate the minimum position of $V$ and the oscillation frequency of the system near the minimum. We start from the many-body Hamiltonian for the three-body recombination process $\mathrm{A}+\mathrm{A}+\mathrm{A}\longleftrightarrow\mathrm{A}_2+\mathrm{A}$ near a Feshbach resonance,
\begin{align}
    \hat{H} = \varepsilon_m\hat{\psi}_m^\dagger\hat{\psi}_m + g_3(\hat{\psi}_m^\dagger\hat{\psi}_a^\dagger\hat{\psi}_a^3 + \hat{\psi}_a^{\dagger3}\hat{\psi}_a\hat{\psi}_m),
\end{align}
where we only consider the ground states for atoms and molecules since their associated coupling terms dominate over others that involve excited states, due to the macroscopic population in the ground state of atoms that we begin with.

To characterize the evolution of the system, we derive the equation of motion for the atomic population $\hat{N}_a = \hat{\psi}_a^\dagger\hat{\psi}_a$ and molecular population $\hat{N}_m = \hat{\psi}_m^\dagger\hat{\psi}_m$, which only depends on the number operators and conserved quantities of the system as~\cite{Barnett1985,Kumar1980}
\begin{align}
    \frac{\hbar^2}{2}\dot{\hat{N}}_m^2 = \frac{\hbar^2}{8}\dot{\hat{N}}_a^2 = -\frac{1}{2}[\hat{N}_m,\hat{H}]^2 = -\frac{1}{2}(\hat{H}-\varepsilon_m\hat{N}_m)^2+g_3^2\hat{N}_a[(\hat{N}_m+1)(\hat{N}_a-1)(\hat{N}_a-2)^2+\hat{N}_m\hat{N}_a(\hat{N}_a+1)(\hat{N}_a+2)],
\end{align}
where $\hat{H}$ is the total energy. We introduce the many-body reaction potential $\hat{V}_3$ through $\hbar^2\dot{\hat{N}}_m^2/2 + \hat{V}(\hat{N}_m) = const.$ as shown in Eq.~(\ref{Conservation}) and obtain it under the boundary condition $\langle\hat{V}_3\rangle|_{ N_m=0,N_a=N}=0$ as:
\begin{align}
  \hat{V}_3 &= \frac{1}{2}[\hat{N}_m,\hat{H}]^2 - \frac{1}{2}[\hat{N}_m,\hat{H}]^2|_{\hat{N}_m=0,\hat{N}_a=\hat{N}}\nonumber\\
  &= -\varepsilon_m\hat{N}_m(\hat{H}-\frac{\varepsilon_m}{2}\hat{N}_m) - g_3^2P_3(\hat{N}-2\hat{N}_m,\hat{N}_m) + g^2\hat{N}(\hat{N}-1)(\hat{N}-2)^2,
\end{align}

\noindent where the three-body polynomial is $P_3(x,y) = x[(y+1)(x-1)(x-2)^2+yx(x+1)(x+2)]$
and $\hat{N}=\hat{N}_a+2\hat{N}_m$ is the conserved total population that commutes with the Hamiltonian. 


\begin{figure}
    \centering
    \includegraphics[width=170mm]{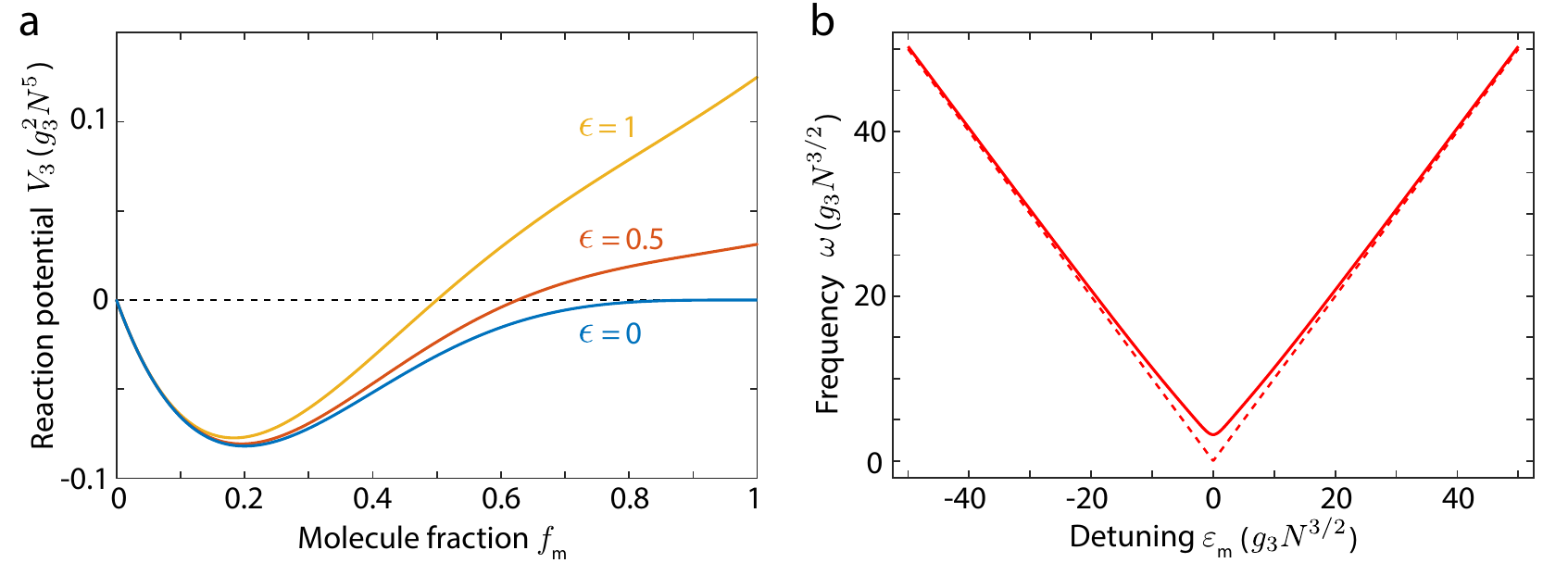}
    \caption{\textbf{The reaction potential $V_3$ for the three-body process and the oscillation frequency $\omega$ near the minimum of $V_3$ at different detuning $\epsilon$.} \textbf{a}, The reaction potential calculated based on Eq.~(\ref{V3_Ninfty}) at $\epsilon = 0$ (blue), $\epsilon = 0.5$ (orange) and $\epsilon = 1$ (yellow). The region below the dashed line at $V_3 = 0$ is where the system is allowed to reach based on the conservation law in Eq.~(\ref{Conservation}). \textbf{b}, The oscillation frequency $\omega$ at different detunings $\varepsilon_m$ (solid line). The dashed line represents the asymptote $|\varepsilon_m|$ in the large detuning limit. On resonance $\epsilon=0$, the frequency is $\omega = 16g_3N^{3/2}/5$.}
    \label{fig:FigS4}
\end{figure}
In our experiment, the atoms and molecules have macroscopic population in the reaction process $N_a,N_m\gg 1$. We may replace all the operators by their expectation values and express the reaction potential in terms of the total particle number $N$ and the molecular fraction $f_m=2N_m/N$ as
\begin{align}
    V_3 &= -g_3^2Nf_m\bigg\{[(1-f_m)^4-\frac{\epsilon^2}{8}f_m]N^4 + (\frac{\epsilon \tilde{H}}{2}+2f_m^3-7f_m^2+9f_m-5)N^3\nonumber\\
    &+5(2f_m^2-5f_m+4)N^2+2(5f_m-9)N+4\bigg\},
\end{align}
where the dimensionless detuning and total energy are $\epsilon = \varepsilon_m/g_3N^{3/2}$ and $\tilde{H} = H/g_3N^{3/2}$, respectively. In the thermodynamic limit where the particle number $N\gg 1$, the potential is reduced to:
\begin{align}
    V_3 = -g_3^2N^5f_m[(1-f_m)^4 - \epsilon^2f_m/8].
    \label{V3_Ninfty}
\end{align}
The potential curves at different detunings are shown in Fig.~\ref{fig:FigS4}a. The potential minimum $f_{m0}$ satisfies $\partial_{f_m}V_3|_{f_{m0}} = 0$. On resonance $\epsilon = 0$, the molecule fraction at the minimum is
\begin{align}
    f_{m0} = \frac{1}{5},
\end{align}
and the oscillation frequency $\omega_0$ of the system around the minimum determined by the curvature of the potential $V_3$ at $f_{m0}$ is
\begin{align}
    \omega_0 = \frac{16}{5}g_3N^{3/2}.
\end{align}
At finite detuning, the minimum position $f_{m0}$ and the oscillation frequency $\omega$ can be solved numerically from $(1-f_{m0})^3(1-5f_{m0}) - \epsilon^2f_{m0}/4 = 0$ and $\omega = g_3N^{3/2}\sqrt{16(1-f_{m0})^2(2-5f_{m0})+\epsilon^2}$, see Fig~\ref{fig:FigS4}b. In the large detuning limit $\epsilon\gg1$, the frequency approaches the absolute value of the molecular energy $|\varepsilon_{m}|$. 

For the nonuniform Thomas-Fermi density distribution of a BEC in a harmonic trap, the coupling strength $g_3$ depends on the particle number and is given by $g_3 = g_3'/\Omega^{3/2}$, where $g_3'$ is a coupling constant, the effective trap volume is $\Omega = (14\pi/15^{2/5})(Na\bar{a}^4)^{3/5}$ and the oscillator length $\bar{a}$ is determined by the trap frequencies as $\bar{a} = \sqrt{\hbar/m(\omega_x\omega_y\omega_z)^{1/3}}$~\cite{Pethick2008}. Thus on resonance the frequency $\omega_0\propto N^{3/5}$. Note that in our experiment, we measure the dependence of the oscillation frequency $\omega_0$ on the initial total atom number $N_0$ and we assume that the atom number in the equilibration phase is proportional to $N_0$. By fitting the data in Fig.~\ref{fig:Fig4}b, we extract the three-body coupling constant $g_3'/h = 5.5(5)\times 10^{-18} \mathrm{cm}^{9/2}/\mathrm{s}$.

Similarly, for the Hamiltonian in Eq.~(\ref{H}) for the two-body process $\mathrm{A} + \mathrm{A}\longleftrightarrow\mathrm{A}_2$, the exact effective potential $\hat{V}_2$ under the boundary condition $\langle\hat{V}_2\rangle|_{ N_m=0,N_a=N}=0$ is derived to be
\begin{align}
    \hat{V}_2 = \varepsilon_m\hat{N}_m(\frac{1}{2}\varepsilon_m\hat{N}_m-\hat{H}) - 2g_2^2\hat{N}_m[(\hat{N}-2\hat{N}_m)^2-\hat{N}+2].
\end{align}
After replacing the operators by their expectation values we obtain
\begin{align}
    V_2 = -g_2^2Nf_m\bigg\{[(1-f_m)^2-\epsilon'^2f_m/8]N^2+(\epsilon'\tilde{H}'/2-1)N+2\bigg\},
\end{align}
where the dimensionless detuning and total energy are $\epsilon' = \varepsilon_m/g_2\sqrt{N}$ and $\tilde{H}' = H/g_2\sqrt{N}$ for the two-body process. In the thermodynamic limit $N\gg 1$, the potential is reduced to:
\begin{align}
    V_2 = -g_2^2N^3f_m[(1-f_m)^2 -\epsilon'^2f_m/8],
    \label{V2_detuning}
\end{align}
based on which we derive the analytical expressions for the minimum position and the oscillation frequency near the minimum as:
\begin{align}
    f_{m0} &= \frac{16+\epsilon'^2-\sqrt{\epsilon'^4+32\epsilon'^2+64}}{24}\nonumber,\\
    \omega &= g_2\sqrt{N}(\epsilon'^4+32\epsilon'^2+64)^{1/4}.
\end{align}
On resonance $\epsilon'=0$, we have 
\begin{align}
    f_{m0} &= \frac{1}{3},\nonumber\\
    \omega_0 &= 2\sqrt{2}g_2\sqrt{N}. 
\end{align}
For the nonuniform density distribution, the frequency is $\omega_0\propto \sqrt{N/\Omega} = N^{1/5}$.

In general, for reactions of order $\alpha$ where $\alpha$ is the maximum number of field operators in the terms of the reaction Hamiltonian, the reaction potential is $V\propto N^\alpha$ in the thermodynamic limit for a uniform system where each mode is macroscopically occupied. Then the oscillation frequency of the system around the potential minimum $f_{m0}$ is determined by the curvature of the potential as $\omega_0 \propto \sqrt{\partial^2 V/\partial N^2|_{f_{m0}}}\propto N^{\alpha/2-1}$.

\section{Evaluation and theory of the molecule formation rate coefficient}
The molecule formation rate coefficient $\beta$ presented in Fig.~\ref{fig:Fig2}b is evaluated based on $\beta = \dot{N}_m/N_0n_0$. Apart from the growth rate of the molecule fraction $f_m = 2N_m/N_0$ shown in Fig.~\ref{fig:Fig2}a, we also need to evaluate the initial mean atomic density $n_0$ to calculate $\beta$. We model the atomic density distribution for the general case where both normal and BEC components exist as~\cite{ketterle1999}
\begin{align}
    n(\bold{r},t) = n_n(\bold{r},t) + n_{\mathrm{B}}(\bold{r},t) =  n_{np}g_{3/2}(\prod_{i=1}^3e^{-r_i^2/\sigma_i^2}) + n_{\mathrm{B}p}(1-\sum_{i=1}^3\frac{r_i^2}{R_i^2})\theta(1-\sum_{i=1}^3\frac{r_i^2}{R_i^2}),
    \label{n_3D}
\end{align}
where $(r_1,r_2,r_3) = (x,y,z)$ are spatial coordinates, $\sigma_i$ and $R_i$ are the widths of the normal and BEC components, $g_{3/2}(x) = \sum_{k=1}^{\infty}x^k/k^{3/2}$ is the polylogarithm function and $\theta(x)$ is the Heaviside step function. Our vertical absorption imaging integrates over the z axis and we obtain the 2D column density as
\begin{align}
    n(x,y,t) = \int_{-\infty}^{\infty}n(\bold{r},t)dz = \sqrt{\pi}n_{np}\sigma_zg_2(\prod_{i=1}^2e^{-r_i^2/\sigma_i^2}) + \frac{4}{3}n_{\mathrm{B}p}R_z(1-\frac{x^2}{R_x^2}-\frac{y^2}{R_y^2})^{3/2}\theta(1-\frac{x^2}{R_x^2}-\frac{y^2}{R_y^2}).
    \label{n_2D}
\end{align}

\begin{figure}
    \centering
    \includegraphics[width=170mm]{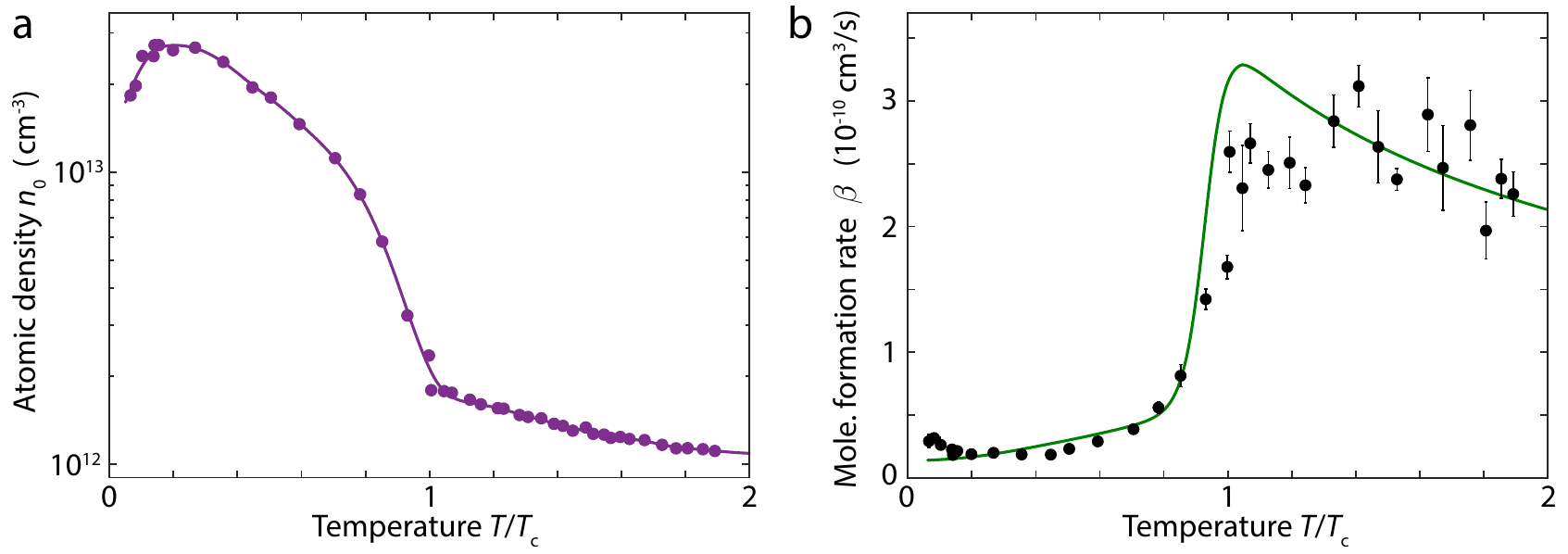}
    \caption{\textbf{Measurement of the mean atomic density $n_0$ and fitting the molecule formation rate coefficient by the two-component model.} \textbf{a}, Initial 3D mean atomic density as a function of temperature, evaluated based on Eq.~(\ref{n_3D_formula}). The solid line is a guide to the eye. \textbf{b}, Molecule formation rate coefficient fit by Eq.~(\ref{beta_mixture}) with the branching ratio $b_{\mathrm{Q}}$ as a fitting parameter (green solid line).}
    \label{fig:FigS3}
\end{figure}

We use Eq.~(\ref{n_2D}) to fit 2D column density distributions from our experiments and extract the widths $\sigma_x$, $\sigma_y$, $R_x$ and $R_y$ and total particle number $N_{n}$ and $N_{\mathrm{B}}$ in the normal and BEC components. We use the scaling to calculate the cloud widths in the z direction as $\sigma_z = (\omega_y/\omega_z)\sigma_y$ and $R_z = (\omega_y/\omega_z)R_y$ based on our independent measurements of trap frequencies $\omega_y$ and $\omega_z$. Then the 3D peak densities are evaluated as
\begin{align}
    n_{np} = \frac{N_n}{g_3(1)\pi^{3/2}\bar{\sigma}^3},~n_{\mathrm{B}p} = \frac{15N_{\mathrm{B}}}{8\pi\bar{R}^3},
\end{align}
where $\bar{\sigma} = (\sigma_x\sigma_y\sigma_z)^{1/3}$ and $\bar{R} = (R_xR_yR_z)^{1/3}$. Eventually, with all the parameters in Eq.~(\ref{n_3D}) determined, we evaluate the initial mean 3D density as
\begin{align}
    n_{0} = \frac{\int n^2(\bold{r},0)d^3\bold{r}}{\int n(\bold{r},0)d^3\bold{r}},
    \label{n_3D_formula}
\end{align}
as shown in Fig.~\ref{fig:FigS3}a.

To capture the sharp transition of the rate coefficient around the critical temperature $T_c$ as shown in Fig.~\ref{fig:Fig2}b, we develop the following two-component model incorporating both the thermal and condensed atoms. There are three possible collision processes leading to molecule formation, which are represented by the following rate equations involving molecular density $m(\bold{r},t)$ and atomic density $n(\bold{r},t)$ with the subscripts $n$ and $\mathrm{B}$ indicating normal and BEC components, respectively.

1. Molecule formation from collisions between condensed atoms:
\begin{equation}
    \dot{m}_\mathrm{B}(\bold{r},t)=\beta_{\mathrm{B}}n^2_\mathrm{B}(\bold{r},t)
    \label{m1}
\end{equation}
with the rate coefficient given by $\beta_{\mathrm{B}} = b_{\mathrm{Q}}\Gamma_{\mathrm{Q}}$, where the collision rate is $\Gamma_{\mathrm{Q}} = \frac{4\sqrt{3}}{(6\pi^5)^{1/3}}\frac{h}{m}\frac{\langle n^{2/3}\rangle}{\langle n\rangle}$, $\langle n\rangle = \int n^2(\textbf{r},t)d^3\textbf{r}/\int n(\textbf{r},t)d^3\textbf{r}$ and $\langle n^{2/3}\rangle = \int n^{5/3}(\textbf{r},t)d^3\textbf{r}/\int n(\textbf{r},t)d^3\textbf{r}$.

2. Molecule formation from collisions between thermal atoms:
\begin{equation}
    \dot{m}_n(\bold{r},t)=\beta_{n}n^2_n(\bold{r},t),
    \label{m2}
\end{equation}
with the rate coefficient given by $\beta_{n} = b_{\mathrm{Cl}}\Gamma_{\mathrm{Cl}}$, where the collision rate is $\Gamma_{\mathrm{Cl}} = 8h^2(\pi^3m^3k_{\mathrm{B}}T)^{-1/2}$.

3. Molecule formation from collisions between one BEC atom and one thermal atom:
\begin{equation}
    \dot{m}_{n\mathrm{B}}(\bold{r},t)=2\beta_{n\mathrm{B}}n_{\mathrm{B}}(\bold{r},t)n_n(\bold{r},t)=2\beta_nn_{\mathrm{B}}(\bold{r},t)n_n(\bold{r},t),
    \label{m3}
\end{equation}
where we assume $\beta_{n\mathrm{B}} = \beta_n = b_{\mathrm{Cl}}\Gamma_{\mathrm{Cl}}$.

The total molecular formation rate is given by the summation of Eqs.~(\ref{m1})-(\ref{m3}):
\begin{align}
    \dot{m}(\textbf{r},t) = \beta_{\mathrm{B}}n_{\mathrm{B}}^2(\textbf{r},t) + \beta_n[n_n(\textbf{r},t)+2n_{\mathrm{B}}(\textbf{r},t)]n_n(\textbf{r},t) = \beta n(\textbf{r},t).
\end{align}
After the spatial integration, we obtain the molecular formation rate coefficient in the mixture
\begin{align}
    \beta = w_{\mathrm{B}}\beta_{\mathrm{B}}+(1 - w_{\mathrm{B}})\beta_n,
    \label{beta_mixture}
\end{align}
where the weight of the BEC component is
\begin{align}
    w_{\mathrm{B}} = \frac{\int n^2_\mathrm{B}(\bold{r},t)d^3\bold{r}}{\int n^2(\bold{r},t)d^3\bold{r}}.
    \label{weights}
\end{align}
For a uniform system, the weight is reduced to $w_{\mathrm{B}} = f^2_{\mathrm{BEC}}$ where $f_{\mathrm{BEC}} = N_{\mathrm{B}}/N$ is the BEC fraction. Since we perform the rate coefficient measurements in a harmonic trap, the density distributions are nonuniform and the weight is enhanced to be $w_{\mathrm{B}} > f^2_{\mathrm{BEC}}$.
Our two-component model captures the transition of the measured rate coefficients around $T_c$ very well, see Fig.~\ref{fig:FigS3}b.

\section{Extraction of molecule oscillation frequency and atom loss rate}
We use the following function to fit the data in the equilibration phase at $1~\mathrm{ms}< t<3~\mathrm{ms}$ shown in Fig.~\ref{fig:Fig3} for the extraction of molecule oscillation frequencies~\cite{Claussen2003}:
\begin{align}
    N_m(t) = N_m(0) - \gamma_1t + \Delta N_me^{-\gamma_2t}\sin(\omega t+\phi),
    \label{fit_Nmt1}
\end{align}
where the fitting parameters are the molecule number $N_m(0)$ extrapolated to time $t=0$, decay rates $\gamma_1$ and $\gamma_2$, oscillation amplitude $\Delta N_m$, oscillation frequency $\omega$ and phase $\phi$. Here the two decay rates $\gamma_1$ and $\gamma_2$ characterize the decay of molecule number and the damping of molecule oscillation amplitude, which are generally different.

For the data shown in Fig.~\ref{fig:Fig4}, we fit the data at $0.3~\mathrm{ms}<t<3~\mathrm{ms}$ using the function
\begin{align}
    N_m(t) = e^{-\gamma t}[N_m(0) + \Delta N_m\sin(\omega t+\phi)],
\end{align}
where we find the single decay rate $\gamma$ is enough to describe the data very well. For each fit, we subtract from the time $t$ a delay time of 0.15 ms due to the finite speed of our magnetic field switch. 
\begin{figure}[h!]
    \centering
    \includegraphics[width=172 mm]{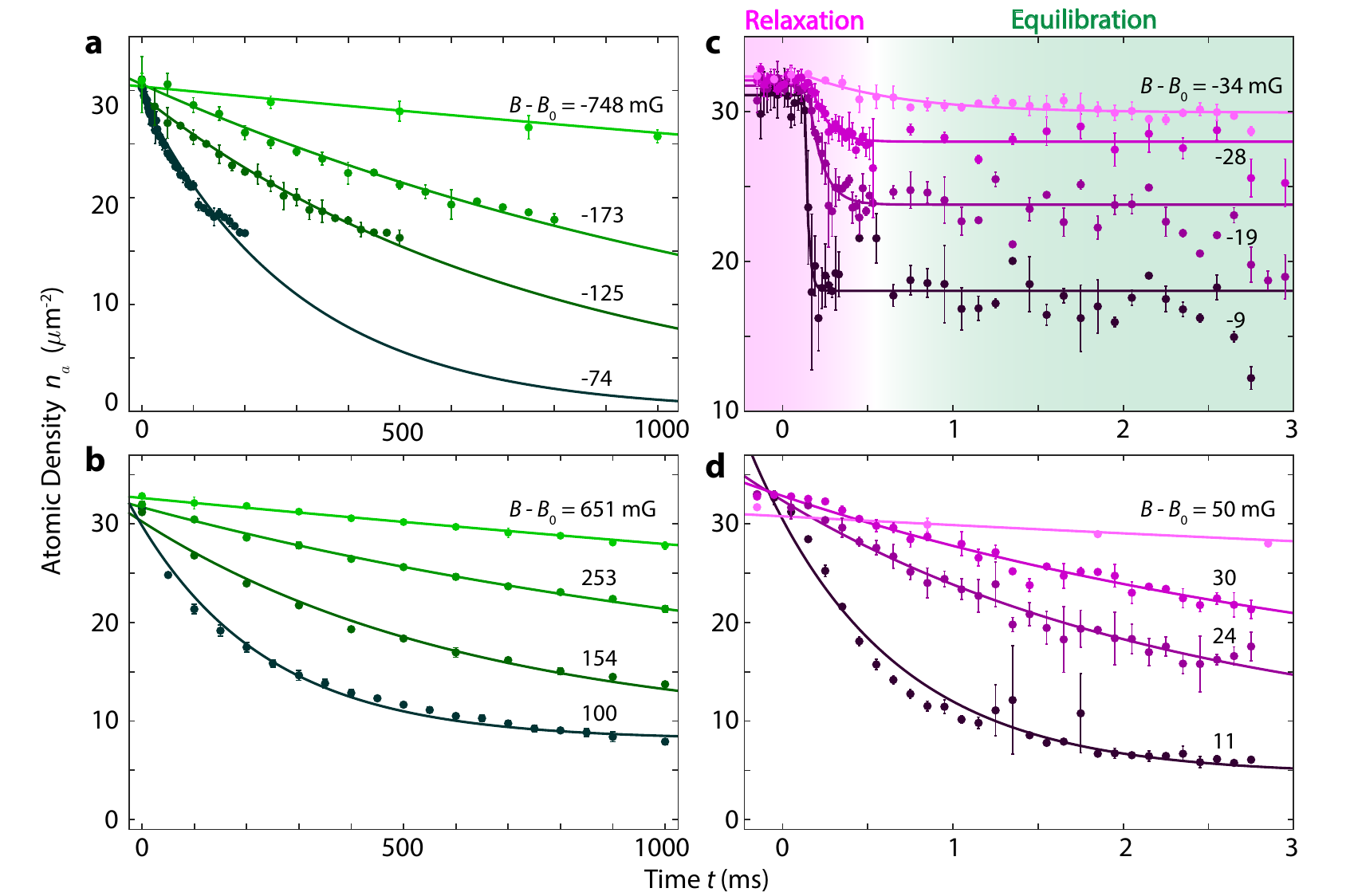}
    \caption{\textbf{Examples of atomic density evolution in a 2D flat-bottomed optical potential for the data presented in Fig.~\ref{fig:Fig3}c.} For data below the resonance, BECs are initially prepared at 19.5~G and magnetic field is quenched to values between 0.05 and 1~G (panel \textbf{a}) and between 5 and 50~mG (panel \textbf{c}) below the resonance. Relaxation and equilibration phases are marked with different background colors in panel \textbf{c}. For data above the resonance, BECs are initially prepared at 20.4~G and magnetic field is quenched to values between 0.1 and 1~G (panel \textbf{b}) and between 10 and 50~mG (panel \textbf{d}) above the resonance. Solid lines are fits for extracting the atom loss rates, see text.}
    \label{fig:FigS6}
\end{figure}

To prevent the fits from getting stuck in a local optimum, we vary the initial guess of the frequency $\omega$ for the fits and use the result that has the minimum root mean square error.

For the atom loss rate measurement shown in Fig.~\ref{fig:Fig3}c, we present example time traces of the averaged atomic density in the 2D flat-bottomed trap in Fig.~\ref{fig:FigS6}. Far from the resonance, see Fig.~\ref{fig:FigS6}a and b, the atomic density decays slowly and we fit the data using

\begin{align}
    n_a(t) = n_a(0)e^{-\gamma_at},
    \label{na_exp}
\end{align}
where $n_a(0)$ is the initial atomic density and $\gamma_a$ is the atom loss rate. The fit is applied to the data above half of the initial density. 

Below and near the resonance, see Fig.~\ref{fig:FigS6}c, we find the density first decays rapidly and then settles around some equilibrium value before a slow decay kicks in at a longer time scale than 3 ms. In this case, we use the following fit function: 
\begin{align}
    n_a(t)=n_a(0)\bigg\{\theta(t_{0}-t)+[(1-s)e^{-\gamma_a(t-t_{0})}+s]\theta(t-t_{0})\bigg\},
\end{align}
where $t_{0}$ is the time when the decay begins and $s$ represents the fractional density the system settles to after the initial fast decay. On the other hand, above and near the resonance, see Fig.~\ref{fig:FigS6}d, the data is fit well by a single exponential decay as in Eq.~(\ref{na_exp}). 

\begin{figure}[h!]
    \centering
    \includegraphics[width=152mm]{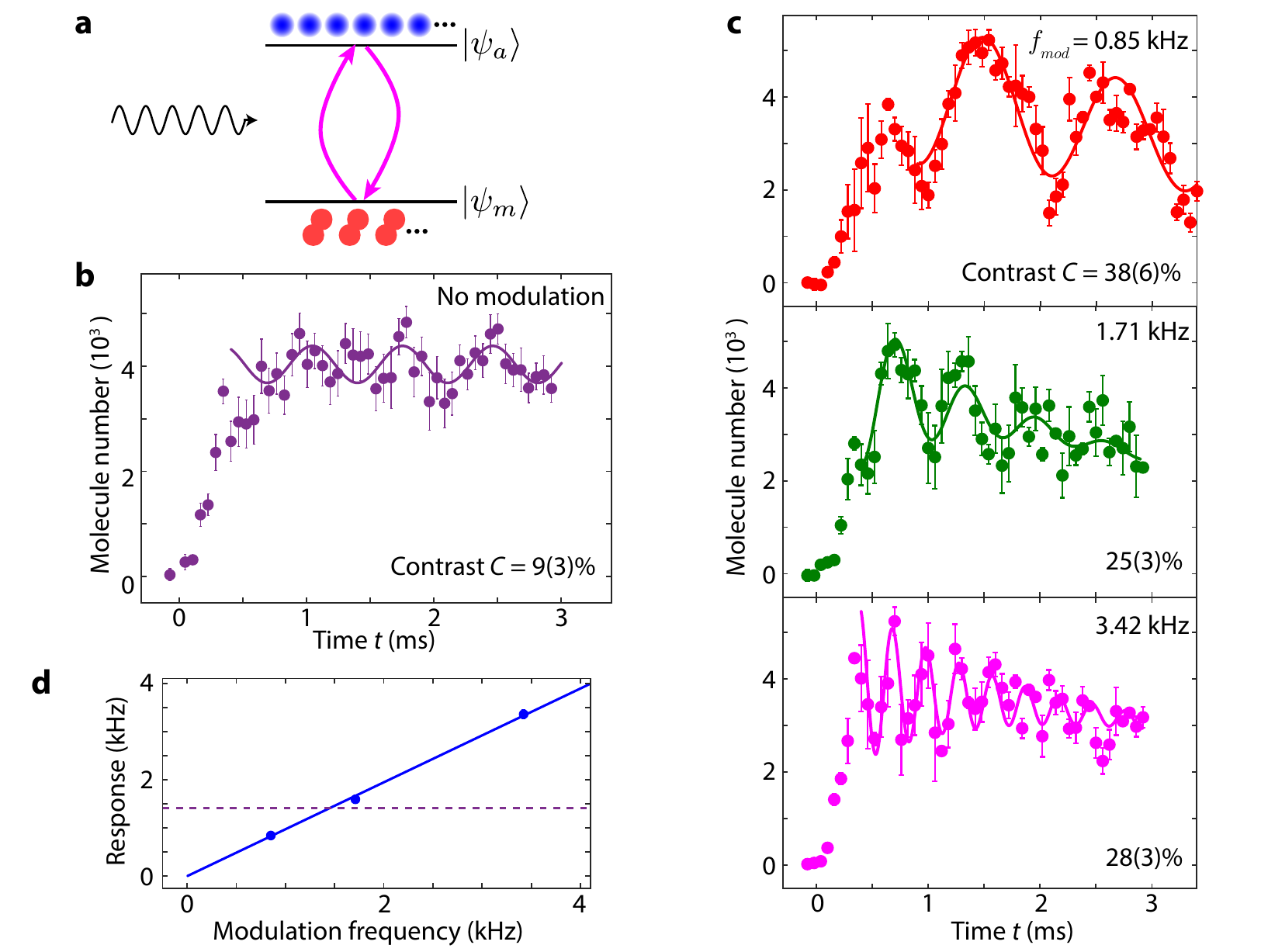}
    \caption{\textbf{Amplification of coherent molecule oscillations through magnetic field modulation.} \textbf{a}, Schematic diagram of molecule formation near the Feshbach resonance with additional sinusoidal magnetic field modulation, where the black wavy arrow represents the RF photon from the modulation. \textbf{b}, Evolution of the molecular population after quench to a static magnetic field 2(1) mG below the resonance, where the contrast $C$ of the molecule oscillation is $9(3)\%$. \textbf{c}, Time traces of molecule number with magnetic field modulation at 0.85 (red), 1.71 (green) and 3.42 kHz (magenta) with modulation amplitude $B_{\mathrm{ac}}\approx 4$ mG around the same field as in panel \textbf{b}. The solid lines are fits based on Eq.~(\ref{fit_Nmt1}). The contrast is defined as $C = N_m(0)/\Delta N_m$, see text. \textbf{d}, Frequency of the molecule oscillations extracted from the fits in panel \textbf{c}. Blue solid line is a linear fit without an offset, which gives a slope 0.97(3). The purple dashed line represents the molecule oscillation frequency in panel \textbf{b}. }
    \label{fig:FigS5}
\end{figure}

\section{Amplification of the molecule oscillation through magnetic field modulation}
We observe large amplitude molecule number oscillations by applying additional external driving to atomic BECs near the Feshbach resonance. The coupled atomic and molecular BECs in our system effectively form a bosonic Josephson junction~\cite{Liu2010}, in close analogy to Cooper-pair Josephson junctions in superconducting devices~\cite{barone1982}. Inspired by the Shapiro resonance effect where adding a small resonant ac component to an applied voltage enhances dc tunneling current in the superconducting Josephson junction, we modulate the magnetic field at a frequency close to the free molecule oscillation frequency at a static field, with the hope to facilitate the reaction. At 2 mG below the resonance, the molecule number oscillates around 4,000 at 1.4 kHz, with a contrast of $\sim$9\% (see Fig.~\ref{fig:FigS5}b). After adding a magnetic field modulation with an amplitude of 4 mG around the same field, we find the oscillation contrast increases by a factor of 3 to 4, see Fig.~\ref{fig:FigS5}c. We also see clear damping in the oscillations with the driving field, likely due to additional heating introduced by the driving.


\begin{thebibliography}{10}
\expandafter\ifx\csname url\endcsname\relax
  \def\url#1{\texttt{#1}}\fi
\expandafter\ifx\csname urlprefix\endcsname\relax\def\urlprefix{URL }\fi
\providecommand{\bibinfo}[2]{#2}
\providecommand{\eprint}[2][]{\url{#2}}

\bibitem{Julienne2006_2}
\bibinfo{author}{K\"ohler, T.}, \bibinfo{author}{G\'oral, K.} \&
  \bibinfo{author}{Julienne, P.~S.}
\newblock \bibinfo{title}{Production of cold molecules via magnetically tunable
  Feshbach resonances}.
\newblock \emph{\bibinfo{journal}{Rev. Mod. Phys.}}
  \textbf{\bibinfo{volume}{78}}, \bibinfo{pages}{1311--1361}
  (\bibinfo{year}{2006}).

\bibitem{Julienne2006_1}
\bibinfo{author}{Jones, K.~M.}, \bibinfo{author}{Tiesinga, E.},
  \bibinfo{author}{Lett, P.~D.} \& \bibinfo{author}{Julienne, P.~S.}
\newblock \bibinfo{title}{Ultracold photoassociation spectroscopy: Long-range
  molecules and atomic scattering}.
\newblock \emph{\bibinfo{journal}{Rev. Mod. Phys.}}
  \textbf{\bibinfo{volume}{78}}, \bibinfo{pages}{483--535}
  (\bibinfo{year}{2006}).

\bibitem{Shuman2010}
\bibinfo{author}{Shuman, E.~S.}, \bibinfo{author}{Barry, J.~F.} \&
  \bibinfo{author}{DeMille, D.}
\newblock \bibinfo{title}{Laser cooling of a diatomic molecule}.
\newblock \emph{\bibinfo{journal}{Nature}} \bibinfo{pages}{820--823}
(\bibinfo{year}{2010}).

\bibitem{Yan2013}
\bibinfo{author}{Yan, B.} \emph{et~al.}
\newblock \bibinfo{title}{Observation of dipolar spin-exchange interactions
  with lattice-confined polar molecules}.
\newblock \emph{\bibinfo{journal}{Nature}} \bibinfo{pages}{521--525}
(\bibinfo{year}{2013}).

\bibitem{Cairncross2021}
\bibinfo{author}{Cairncross, W.~B.} \emph{et~al.}
\newblock \bibinfo{title}{Assembly of a rovibrational ground state molecule in
  an optical tweezer}.
\newblock \emph{\bibinfo{journal}{Phys. Rev. Lett.}}
  \textbf{\bibinfo{volume}{126}}, \bibinfo{pages}{123402}
  (\bibinfo{year}{2021}).

\bibitem{Joschka2017}
\bibinfo{author}{Wolf, J.} \emph{et~al.}
\newblock \bibinfo{title}{State-to-state chemistry for three-body recombination
  in an ultracold rubidium gas}.
\newblock \emph{\bibinfo{journal}{Science}} \bibinfo{pages}{921--924}
(\bibinfo{year}{2017}).

\bibitem{Rui2017}
\bibinfo{author}{Rui, J.} \emph{et~al.}
\newblock \bibinfo{title}{Controlled state-to-state atom-exchange reaction in
  an ultracold atom--dimer mixture}.
\newblock \emph{\bibinfo{journal}{Nature Physics}} \bibinfo{pages}{699--703}
(\bibinfo{year}{2017}).

\bibitem{Liu2021}
\bibinfo{author}{Liu, Y.} \emph{et~al.}
\newblock \bibinfo{title}{Precision test of statistical dynamics with
  state-to-state ultracold chemistry}.
\newblock \emph{\bibinfo{journal}{Nature}} \bibinfo{pages}{379--384}
(\bibinfo{year}{2021}).

\bibitem{Qijin2005}
\bibinfo{author}{Chen, Q.}, \bibinfo{author}{Stajic, J.}, \bibinfo{author}{Tan,
  S.} \& \bibinfo{author}{Levin, K.}
\newblock \bibinfo{title}{BCS--BEC crossover: From high temperature
  superconductors to ultracold superfluids}.
\newblock \emph{\bibinfo{journal}{Physics Reports}} \bibinfo{pages}{1--88}
(\bibinfo{year}{2005}).

\bibitem{Giorgini2008}
\bibinfo{author}{Giorgini, S.}, \bibinfo{author}{Pitaevskii, L.~P.} \&
  \bibinfo{author}{Stringari, S.}
\newblock \bibinfo{title}{Theory of ultracold atomic Fermi gases}.
\newblock \emph{\bibinfo{journal}{Rev. Mod. Phys.}}
  \textbf{\bibinfo{volume}{80}}, \bibinfo{pages}{1215--1274}
  (\bibinfo{year}{2008}).

\bibitem{Marco2019}
\bibinfo{author}{Marco, L.~D.} \emph{et~al.}
\newblock \bibinfo{title}{A degenerate Fermi gas of polar molecules}.
\newblock \emph{\bibinfo{journal}{Science}} \bibinfo{pages}{853--856}
(\bibinfo{year}{2019}).

\bibitem{Zhang2021}
\bibinfo{author}{Zhang, Z.}, \bibinfo{author}{Chen, L.}, \bibinfo{author}{Yao,
  K.-X.} \& \bibinfo{author}{Chin, C.}
\newblock \bibinfo{title}{Transition from an atomic to a molecular
  Bose--Einstein condensate}.
\newblock \emph{\bibinfo{journal}{Nature}} \bibinfo{pages}{708--711}
(\bibinfo{year}{2021}).

\bibitem{Heinzen2000}
\bibinfo{author}{Heinzen, D.~J.}, \bibinfo{author}{Wynar, R.},
  \bibinfo{author}{Drummond, P.~D.} \& \bibinfo{author}{Kheruntsyan, K.~V.}
\newblock \bibinfo{title}{Superchemistry: Dynamics of coupled atomic and
  molecular Bose-Einstein condensates}.
\newblock \emph{\bibinfo{journal}{Phys. Rev. Lett.}}
  \textbf{\bibinfo{volume}{84}}, \bibinfo{pages}{5029--5033}
  (\bibinfo{year}{2000}).

\bibitem{Malla2021}
\bibinfo{author}{Malla, R.~K.}, \bibinfo{author}{Chernyak, V.~Y.},
  \bibinfo{author}{Sun, C.} \& \bibinfo{author}{Sinitsyn, N.~A.}
\newblock \bibinfo{title}{Coherent reaction between molecular and atomic
  Bose-Einstein condensates: Integrable model}.
\newblock \emph{\bibinfo{journal}{Phys. Rev. Lett.}}
  \textbf{\bibinfo{volume}{129}}, \bibinfo{pages}{033201}
  (\bibinfo{year}{2022}).

\bibitem{Moore2002}
\bibinfo{author}{Moore, M.~G.} \& \bibinfo{author}{Vardi, A.}
\newblock \bibinfo{title}{Bose-enhanced chemistry: Amplification of selectivity
  in the dissociation of molecular Bose-Einstein condensates}.
\newblock \emph{\bibinfo{journal}{Phys. Rev. Lett.}}
  \textbf{\bibinfo{volume}{88}}, \bibinfo{pages}{160402}
  (\bibinfo{year}{2002}).

\bibitem{Vardi2001}
\bibinfo{author}{Vardi, A.}, \bibinfo{author}{Yurovsky, V.~A.} \&
  \bibinfo{author}{Anglin, J.~R.}
\newblock \bibinfo{title}{Quantum effects on the dynamics of a two-mode
  atom-molecule Bose-Einstein condensate}.
\newblock \emph{\bibinfo{journal}{Phys. Rev. A}} \textbf{\bibinfo{volume}{64}},
  \bibinfo{pages}{063611} (\bibinfo{year}{2001}).

\bibitem{Richter_2015}
\bibinfo{author}{Richter, F.} \emph{et~al.}
\newblock \bibinfo{title}{Ultracold chemistry and its reaction kinetics}.
\newblock \emph{\bibinfo{journal}{New Journal of Physics}}
  \bibinfo{pages}{055005}
  (\bibinfo{year}{2015}).


\bibitem{Clark2017}
\bibinfo{author}{Clark, L.~W.}, \bibinfo{author}{Gaj, A.},
  \bibinfo{author}{Feng, L.} \& \bibinfo{author}{Chin, C.}
\newblock \bibinfo{title}{Collective emission of matter-wave jets from driven
  Bose--Einstein condensates}.
\newblock \emph{\bibinfo{journal}{Nature}} \bibinfo{pages}{356--359}
(\bibinfo{year}{2017}).

\bibitem{Chin2010}
\bibinfo{author}{Chin, C.}, \bibinfo{author}{Grimm, R.},
  \bibinfo{author}{Julienne, P.} \& \bibinfo{author}{Tiesinga, E.}
\newblock \bibinfo{title}{Feshbach resonances in ultracold gases}.
\newblock \emph{\bibinfo{journal}{Rev. Mod. Phys.}}
  \textbf{\bibinfo{volume}{82}}, \bibinfo{pages}{1225--1286}
  (\bibinfo{year}{2010}).

\bibitem{Greene2017}
\bibinfo{author}{Greene, C.~H.}, \bibinfo{author}{Giannakeas, P.} \&
  \bibinfo{author}{P\'erez-R\'{\i}os, J.}
\newblock \bibinfo{title}{Universal few-body physics and cluster formation}.
\newblock \emph{\bibinfo{journal}{Rev. Mod. Phys.}}
  \textbf{\bibinfo{volume}{89}}, \bibinfo{pages}{035006}
  (\bibinfo{year}{2017}).

\bibitem{Jin2014}
\bibinfo{author}{Makotyn, P.}, \bibinfo{author}{Klauss, C.~E.},
  \bibinfo{author}{Goldberger, D.~L.}, \bibinfo{author}{Cornell, E.~A.} \&
  \bibinfo{author}{Jin, D.~S.}
\newblock \bibinfo{title}{Universal dynamics of a degenerate unitary Bose gas}.
\newblock \emph{\bibinfo{journal}{Nature Physics}} \bibinfo{pages}{116--119}
(\bibinfo{year}{2014}).

\bibitem{Eismann2016}
\bibinfo{author}{Eismann, U.} \emph{et~al.}
\newblock \bibinfo{title}{Universal loss dynamics in a unitary Bose gas}.
\newblock \emph{\bibinfo{journal}{Phys. Rev. X}} \textbf{\bibinfo{volume}{6}},
  \bibinfo{pages}{021025} (\bibinfo{year}{2016}).

\bibitem{Petrov2004}
\bibinfo{author}{Petrov, D.~S.}
\newblock \bibinfo{title}{Three-boson problem near a narrow Feshbach
  resonance}.
\newblock \emph{\bibinfo{journal}{Phys. Rev. Lett.}}
  \textbf{\bibinfo{volume}{93}}, \bibinfo{pages}{143201}
  (\bibinfo{year}{2004}).

\bibitem{Chin2004}
\bibinfo{author}{Chin, C.} \& \bibinfo{author}{Grimm, R.}
\newblock \bibinfo{title}{Thermal equilibrium and efficient evaporation of an
  ultracold atom-molecule mixture}.
\newblock \emph{\bibinfo{journal}{Phys. Rev. A}} \textbf{\bibinfo{volume}{69}},
  \bibinfo{pages}{033612} (\bibinfo{year}{2004}).

\bibitem{Pethick2008}
\bibinfo{author}{Pethick, C.~J.} \& \bibinfo{author}{Smith, H.}
\newblock \emph{\bibinfo{title}{Bose--Einstein condensation in dilute gases}}
  (\bibinfo{publisher}{Cambridge university press}, \bibinfo{year}{2008}).

\bibitem{Hung2008}
\bibinfo{author}{Hung, C.-L.}, \bibinfo{author}{Zhang, X.},
  \bibinfo{author}{Gemelke, N.} \& \bibinfo{author}{Chin, C.}
\newblock \bibinfo{title}{Accelerating evaporative cooling of atoms into
  Bose-Einstein condensation in optical traps}.
\newblock \emph{\bibinfo{journal}{Phys. Rev. A}} \textbf{\bibinfo{volume}{78}},
  \bibinfo{pages}{011604} (\bibinfo{year}{2008}).

\bibitem{Chin2005}
\bibinfo{author}{Chin, C.} \emph{et~al.}
\newblock \bibinfo{title}{Observation of Feshbach-like resonances in collisions
  between ultracold molecules}.
\newblock \emph{\bibinfo{journal}{Phys. Rev. Lett.}}
  \textbf{\bibinfo{volume}{94}}, \bibinfo{pages}{123201}
  (\bibinfo{year}{2005}).


\bibitem{zhang2022coherent}
\bibinfo{author}{Zhang, Z.}
\newblock \emph{\bibinfo{title}{Coherent dynamics and reactions in atomic and
  molecular Bose-Einstein condensates}} (\bibinfo{publisher}{The University of Chicago},
  \bibinfo{year}{2022}).

\bibitem{Thompson2005}
\bibinfo{author}{Thompson, S.~T.}, \bibinfo{author}{Hodby, E.} \&
  \bibinfo{author}{Wieman, C.~E.}
\newblock \bibinfo{title}{Ultracold molecule production via a resonant
  oscillating magnetic field}.
\newblock \emph{\bibinfo{journal}{Phys. Rev. Lett.}}
  \textbf{\bibinfo{volume}{95}}, \bibinfo{pages}{190404}
  (\bibinfo{year}{2005}).

\bibitem{Lange2009}
\bibinfo{author}{Lange, A.~D.} \emph{et~al.}
\newblock \bibinfo{title}{Determination of atomic scattering lengths from
  measurements of molecular binding energies near Feshbach resonances}.
\newblock \emph{\bibinfo{journal}{Phys. Rev. A}} \textbf{\bibinfo{volume}{79}},
  \bibinfo{pages}{013622} (\bibinfo{year}{2009}).

\bibitem{hung2011situ}
\bibinfo{author}{Hung, C.-L.}
\newblock \emph{\bibinfo{title}{In situ probing of two-dimensional quantum
  gases}} (\bibinfo{publisher}{The University of Chicago},
  \bibinfo{year}{2011}).

\bibitem{Hung2011}
\bibinfo{author}{Hung, C.-L.}, \bibinfo{author}{Zhang, X.},
  \bibinfo{author}{Gemelke, N.} \& \bibinfo{author}{Chin, C.}
\newblock \bibinfo{title}{Observation of scale invariance and universality in
  two-dimensional Bose gases}.
\newblock \emph{\bibinfo{journal}{Nature}} \bibinfo{pages}{236--239}
  \bibinfo{year}{2011}).

\bibitem{Castin1996}
\bibinfo{author}{Castin, Y.} \& \bibinfo{author}{Dum, R.}
\newblock \bibinfo{title}{Bose-Einstein condensates in time dependent traps}.
\newblock \emph{\bibinfo{journal}{Phys. Rev. Lett.}}
  \textbf{\bibinfo{volume}{77}}, \bibinfo{pages}{5315--5319}
  (\bibinfo{year}{1996}).

\bibitem{Berninger2013}
\bibinfo{author}{Berninger, M.} \emph{et~al.}
\newblock \bibinfo{title}{Feshbach resonances, weakly bound molecular states,
  and coupled-channel potentials for cesium at high magnetic fields}.
\newblock \emph{\bibinfo{journal}{Phys. Rev. A}} \textbf{\bibinfo{volume}{87}},
  \bibinfo{pages}{032517} (\bibinfo{year}{2013}).

\bibitem{Mark2018}
\bibinfo{author}{Mark, M.~J.}, \bibinfo{author}{Meinert, F.},
  \bibinfo{author}{Lauber, K.} \& \bibinfo{author}{Nagerl, H.-C.}
\newblock \bibinfo{title}{{Mott-insulator-aided detection of ultra-narrow
  Feshbach resonances}}.
\newblock \emph{\bibinfo{journal}{SciPost Phys.}} \textbf{\bibinfo{volume}{5}},
  \bibinfo{pages}{55} (\bibinfo{year}{2018}).

\bibitem{Barnett1985}
\bibinfo{author}{Barnett, S.~M.} \& \bibinfo{author}{Knight, P.~L.}
\newblock \bibinfo{title}{Thermofield analysis of squeezing and statistical
  mixtures in quantum optics}.
\newblock \emph{\bibinfo{journal}{J. Opt. Soc. Am. B}}
  \bibinfo{pages}{467--479}
   (\bibinfo{year}{1985}).

\bibitem{Kumar1980}
\bibinfo{author}{Kumar, S.} \& \bibinfo{author}{Mehta, C.~L.}
\newblock \bibinfo{title}{Theory of the interaction of a single-mode resonant
  radiation field with $n$ two-level atoms}.
\newblock \emph{\bibinfo{journal}{Phys. Rev. A}} \textbf{\bibinfo{volume}{21}},
  \bibinfo{pages}{1573--1588} (\bibinfo{year}{1980}).

\bibitem{ketterle1999}
\bibinfo{author}{Ketterle, W.}, \bibinfo{author}{Durfee, D.~S.} \&
  \bibinfo{author}{Stamper-Kurn, D.}
\newblock \bibinfo{title}{Making, probing and understanding Bose-Einstein
  condensates}.
\newblock \emph{\bibinfo{journal}{arXiv preprint cond-mat/9904034}}
  (\bibinfo{year}{1999}).

\bibitem{Claussen2003}
\bibinfo{author}{Claussen, N.~R.} \emph{et~al.}
\newblock \bibinfo{title}{Very-high-precision bound-state spectroscopy near a
  ${}^{85}\mathrm{Rb}$ Feshbach resonance}.
\newblock \emph{\bibinfo{journal}{Phys. Rev. A}} \textbf{\bibinfo{volume}{67}},
  \bibinfo{pages}{060701} (\bibinfo{year}{2003}).

\bibitem{Liu2010}
\bibinfo{author}{Liu, B.}, \bibinfo{author}{Fu, L.-B.} \& \bibinfo{author}{Liu,
  J.}
\newblock \bibinfo{title}{Shapiro-like resonance in ultracold molecule
  production via an oscillating magnetic field}.
\newblock \emph{\bibinfo{journal}{Phys. Rev. A}} \textbf{\bibinfo{volume}{81}},
  \bibinfo{pages}{013602} (\bibinfo{year}{2010}).

\bibitem{barone1982}
\bibinfo{author}{Barone, A.} \& \bibinfo{author}{Paterno, G.}
\newblock \emph{\bibinfo{title}{Physics and applications of the Josephson
  effect}}, vol.~\bibinfo{volume}{1} (\bibinfo{publisher}{Wiley Online
  Library}, \bibinfo{year}{1982}).

\end{thebibliography}
\end{document}